\documentclass[conference]{IEEEtran}

\usepackage{pdflscape}
\usepackage{mathtools}
\usepackage{subcaption}
\usepackage{xcolor}
\usepackage{amsthm}
\usepackage{enumitem}
\usepackage{algorithm}
\usepackage{algpseudocode}
\usepackage{url}
\usepackage{multirow}
\usepackage{tabularx} 
\usepackage{ragged2e} 
\usepackage{diagbox}
\usepackage{xcolor}
\usepackage{colortbl}

\definecolor{lightgreen}{rgb}{0.9,1.0,0.9} 

\newtheorem{definition}{Definition}
\newcommand\TILDE{\char`\~}

\ifCLASSOPTIONcompsoc
  \usepackage[nocompress]{cite}
\else
  \usepackage{cite}
\fi
\ifCLASSINFOpdf

\begin{document}
%
\title{WiFinger: Fingerprinting Noisy IoT Event Traffic Using Packet-level Sequence Matching}

\author{%

    \IEEEauthorblockN{%
        Ronghua Li\IEEEauthorrefmark{1},
        Shinan Liu\IEEEauthorrefmark{2},
        Haibo Hu\IEEEauthorrefmark{1},%
        Qingqing Ye\IEEEauthorrefmark{1},%
        Nick Feamster\IEEEauthorrefmark{3}%
    }%
    
    \IEEEauthorblockA{%
        \IEEEauthorrefmark{1}The Hong Kong Polytechnic University 
        \IEEEauthorrefmark{2}The University of Hong Kong \IEEEauthorrefmark{3}University of Chicago\\
    }%
    \IEEEauthorblockA{%
        \IEEEauthorrefmark{1}cory-ronghua.li@connect.polyu.hk,\{haibo.hu, qqing.ye\}@polyu.edu.hk\\
        \IEEEauthorrefmark{2}shinan6@hku.edu.hk \IEEEauthorrefmark{3}feamster@uchicago.edu
    }%
}

\IEEEoverridecommandlockouts
\makeatletter\def\@IEEEpubidpullup{6.5\baselineskip}\makeatother
\IEEEpubid{\parbox{\columnwidth}{
		Network and Distributed System Security (NDSS) Symposium 2025\\
		24-28 February 2025, San Diego, CA, USA\\
		ISBN 979-8-9894372-8-3\\
		https://dx.doi.org/10.14722/ndss.2025.[23$|$24]xxxx\\
		www.ndss-symposium.org
}
\hspace{\columnsep}\makebox[\columnwidth]{}}

\maketitle

\begin{abstract}
  IoT environments such as smart homes are susceptible to privacy inference attacks, where attackers can analyze patterns of encrypted network traffic to infer the state of devices and even the activities of people. While most existing attacks exploit ML techniques for discovering such traffic patterns, they underperform on wireless traffic, especially Wi-Fi, due to its heavy noisiness and the packet loss of wireless sniffing. In addition, these approaches commonly target distinguishing chunked IoT event traffic samples, and they fail at effectively tracking multiple events simultaneously. In this work, we propose WiFinger, a fine-grained multi-IoT event fingerprinting approach against noisy traffic. WiFinger turns the traffic pattern classification task into a subsequence matching problem and introduces novel techniques to account for the high time complexity while maintaining high accuracy. In addition, its reliance on training sample volumes reduces efforts for any future fingerprint updates. Experiments demonstrate that WiFinger outperforms existing approaches under practical threat models, with an average recall of 89\% (v.s. 49\% and 46\% respectively) and almost zero false positives for various IoT events.
\end{abstract}

\IEEEpeerreviewmaketitle

\section{Introduction}\label{sec:intro}
IoT devices are increasingly ubiquitous in various applications, including smart homes, smart cities, and industrial automation. These devices connect to the internet for control and to transmit sensor data. However, even if the data transmitted is encrypted to prevent information leakage, existing works have demonstrated that device activities can still be inferred and exploited by passively monitoring the encrypted traffic and analyzing the patterns of traffic flows or packets~\cite{pingpong2019trimananda,ren2019information,copos2016anybody,alrawi2019sok}. Consequently, this may expose critical information, such as user behavior or the working status of security-sensitive devices, posing potential security risks for malicious actors (e.g., break-in) or leading to privacy breaches through unauthorized surveillance.

Currently, most existing benign or malicious IoT event/device fingerprinting methods~\cite{pingpong2019trimananda,wan2021iotathena,homesnitch,ren2019information,copos2016anybody,apthorpe2017spying} focus on wired traffic at the TCP/IP layers, with only a few specifically studying fingerprinting on Wi-Fi traffic~\cite{peekaboo,zou2023iotbeholder}. While TCP/IP headers provide verbose information assisting accurate fingerprinting, TCP/IP sniffing poses stringent requirement on attackers~\cite{li2022foap}, who must either exploit the LAN access point or obtain authorized access within the ISP's network (WAN). In contrast, since most IoT devices connect wirelessly (Wi-Fi in this work), their traffic can be easily sniffed using a wireless adapter, significantly lowering the barrier and increasing the practicality of such attacks.

Given that network event classification has been studied for decades, an intuitive question arises: can these established solutions simply work for Wi-Fi layer attacks? Unfortunately, our findings indicate the opposite. By applying prevalent methods to IoT traffic at the Wi-Fi layer, we have identified significant limitations and challenges of these solutions in tuning model performance and reducing manual efforts, as detailed in Table~\ref{tab:comparison}.

\begin{table}[h!]
\centering
\footnotesize
\begin{tabularx}{\linewidth}{|p{1.75cm}|X|p{.6cm}|p{.6cm}|p{.9cm}|p{1.1cm}|}
\hline
\textbf{Methods}& \textbf{Features}  & \textbf{Data Vol} & \textbf{Label Acc} & \textbf{Param Tuning} & \textbf{Traffic Completeness} \\ \hline
DL \cite{aceto2019mimetic,lotfollahi2020deep,rezaei2019large}& Raw Traffic & \textcolor{red}{$\uparrow$} & \textcolor{red}{$\uparrow$} & \textcolor{red}{$\uparrow$} & $-$ \\ \hline
ML \cite{sivanathan2018classifying,iotsentinel,dong2020lstmfinger,ma2021inferringspatial,peekaboo,zou2023iotbeholder,alyami2022wifi}& Traffic stats & $\downarrow$ & \textcolor{red}{$\uparrow$}  & \textcolor{red}{$\uparrow$} & $-$ \\ \hline
Packet Match \cite{pingpong2019trimananda,wan2021iotathena} & Size \& Dir \& Time & $\downarrow$ & $\downarrow$ & $\downarrow$ & \textcolor{red}{$\uparrow$} \\ \hline
\rowcolor{lightgreen}
\textbf{WiFinger} & Size \& Dir \& Time & $\downarrow$ & $\downarrow$ & $\downarrow$ & $\downarrow$ \\ \hline
\end{tabularx}
\caption{Factors that influence the performance of existing approaches. $\uparrow$ indicates high dependence of performances on the factor, and vice versa. $-$ indicates a certain level of robustness against a factor. The lower reliance, the higher feasibility and scalability.}
\label{tab:comparison}
\end{table}

First, the most widely adopted ML/DL-based traffic analysis approaches face limitations in adaptability~\cite{aceto2019mimetic,lotfollahi2020deep,rezaei2019large,peekaboo}, granularity~\cite{peekaboo,saidi2020haystack,nprint}, and scalability for effective IoT event fingerprinting. \textbf{Adaptability:} These methods typically rely on handcrafted features or parameters (e.g., packet size statistics, flow duration) to classify traffic bursts or flows. Nonetheless, the most suitable feature/parameter sets vary among devices and events due to the sparseness and diversity of IoT traffic~\cite{evidence}. \textbf{Granularity:} ML-based methods are not fine-grained enough for event identification, particularly when differentiating events with only slight variations in packet sizes or intervals. Their results can also be easily affected by heartbeat or other background traffic, which influences flow characteristics. \textbf{Scalability:} Training these data-driven models requires a large amount of labeled data from scratch, which does not scale well for large IoT systems.

Second, packet-matching approaches~\cite{pingpong2019trimananda,wan2021iotathena}, albeit accurate, encounter the following two challenges on Wi-Fi traffic. \textbf{Packet Losses:} Passive Wi-Fi sniffing is susceptible to packet loss due to variable wireless channel conditions. This affects both online detection and offline training of packet-matching approaches. \textbf{Upper-layer Variances:} Wi-Fi traffic inherits variances from upper layers, introducing significant noise. For example, Ping-Pong~\cite{pingpong2019trimananda} filtered retransmission/ACK TCP packets and only focused on fingerprint TLS application packets. However, all of them become indistinguishable at the Wi-Fi layer due to WPA encryption~\cite{wpa}, hindering packet-matching effectiveness.

This work aims to accurately fingerprint Wi-Fi IoT events by addressing the above limitations and challenges. To this end, we propose an intuitive and yet noise-agnostic packet-matching approach for wireless traffic, \textit{WiFinger}. Inspired by the findings in \cite{pingpong2019trimananda,wan2021iotathena}, WiFinger adopts a similar fingerprint representation: a sequence of packets with \textit{relative timestamps}, \textit{sizes}, and \textit{directions}. Yet, instead of finding exact match, WiFinger focuses on detecting whether ``traces'' of fingerprint are present within the examined sequence. WiFinger offers three main characteristics: \textbf{(i) Adaptable:} WiFinger uses a consistent parameter setting for fingerprint extraction and matching, requiring little parameter tuning for various devices and events. \textbf{(ii) Accurate:} It is robust against Wi-Fi traffic noise and variances, and can differentiate subtle differences in packet size and inter-arrival time features distinguishing different events. Evaluation on 31 real IoT events shows that WiFinger achieves excellent performance on continuous event tracking with almost zero false positives, even including complex events like voice commands of smart speakers. \textbf{(iii) Scalable:} WiFinger fingerprints can be extracted efficiently using a few dozen of event samples, significantly reducing data collection and model update costs. In summary, our main contributions are as follows:

\begin{itemize}
    \item We provide insights into the differences between the Wi-Fi and TCP/IP traffic, revealing the main challenges of fingerprinting IoT event under wireless IoT traffic.
    \item We identify the current gap between experimental settings and online settings, showing that the performance of existing models are often exaggerated and unrealistic.
    \item We propose WiFinger, a fine-grained IoT event fingerprinting approach on Wi-Fi traffic, formulating the detection as a subsequence matching problem. We tackle the NP-hard efficiency challenge to ensure fast offline training and online matching while maintaining high accuracy. Moreover, we overcome the challenge of extracting fingerprints from noisy traffic using collective intelligence.
\end{itemize}

The rest of the paper is organized as follows. Section~\ref{sec:related} uses real examples to show the limitations of existing methods and motivate this work. Section~\ref{sec:problem} defines the problem scope and threat model. Section~\ref{sec:wifinger} introduces the intuition behind WiFinger and details its fingerprint extraction and matching process. Section~\ref{sec:experiment} compares WiFinger to two state-of-the-art methods and investigates its performance against defenses. Section~\ref{sec:discussion} provides additional findings on IoT event traffic and discusses how WiFinger can be extended to other domains. 
\section{Related Work \& Motivation}\label{sec:related}
Traffic analysis and event detection have been widely studied over the last decade~\cite{pingpong2019trimananda,peekaboo,homesnitch,apthorpe2018keepingprivate,ren2019information,wan2021iotathena,copos2016anybody,iotinspector}.
However, most existing works struggle with fingerprinting Wi-Fi IoT events effectively, particularly compared to methods at higher layers or different protocols.
This section uses real Wi-Fi event sequences to illustrate Wi-Fi IoT traffic characteristics, highlighting limitations and challenges for existing fingerprinting methods in this domain.

\subsection{Wi-Fi Event Traffic: A Motivating Example}
\label{sec:example_fingerprint}
\begin{figure}[h!] 
\centering
\includegraphics[width=\linewidth]{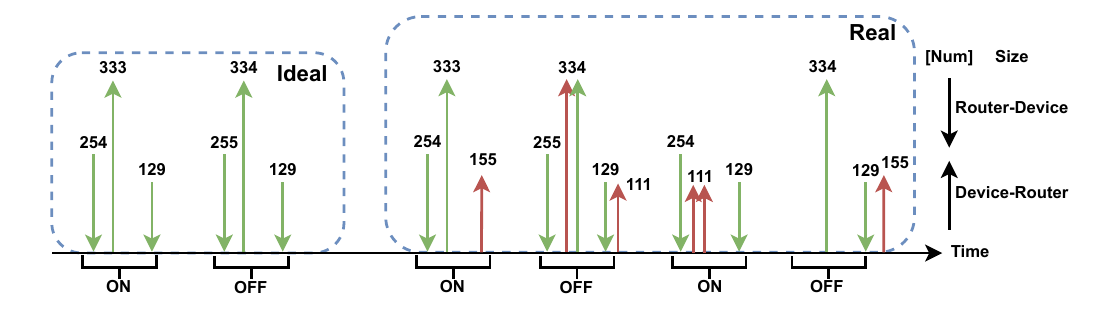}
\caption{An example of ideal Wi-Fi IoT event traffic v.s. real-world collected traffic. Green arrows represent fingerprint packets, and red arrows represent unrelated packets. While the ideal event fingerprints are clean, real-world event traffic may be incomplete or mixed with noise packets.}
\label{fig:example_fingerprint}
\end{figure}

Using a real example, we demonstrate Wi-Fi IoT event traffic characteristics and their differences from typical TCP/IP traffic. Since Wi-Fi Control and Management packets are not related to application-layer behaviors, we only analyze Data packets with payload, similar to \cite{pingpong2019trimananda} on TCP/IP traffic. Figure~\ref{fig:example_fingerprint} shows example Wi-Fi Data packet sequences from a Hue Light Bulb. Down arrows ($D_{size}$) represent downstream packets (router to device) and up arrows ($U_{size}$) represent upstream packets (device to router). Ideally, ``ON'' commands start with a $D_{254}$ packet, followed by a $U_{333}$ upstream packet 0.1 - 0.2s later, and ends with a $D_{129}$ packet after another 0.4s. ``OFF'' commands are similar to ``ON'', except their first two packets are one byte larger.

\smallskip
\noindent \textbf{INSIGHT \#1: Wi-Fi traffic has severe data packet loss due to the nature of wireless sniffing.} Though Wi-Fi includes retransmissions for station-to-AP reliability, they don't guarantee completeness for a passive sniffer in promiscuous mode. Compared to ideal event traffic, real event traffic may have missing components due to the unreliable Wi-Fi sniffing. In contrast, TCP/IP traffic sniffing via port mirroring is typically less susceptible to packet loss. According to our experiments, the best performing adapter suffers from 5-20\% data packet drop, which significantly impacts the detecting performance.

\smallskip
\noindent \textbf{INSIGHT \#2: Event bursts/flows can be dominated by noise packets, obscuring actual event patterns.} As illustrated in Figure~\ref{fig:example_fingerprint}, observed 'Real' sequences often deviate from idealized patterns, being interleaved with packets that are irrelevant with application behaviors. In contrast to TCP/IP payload analysis, Wi-Fi frames also encapsulate transport layer control packets (e.g., ACKs, SYNs), which are not directly part of the application data and ideally should be filtered. Furthermore, when IoT events occur concurrently with other background communication, their simple flow patterns will be mixed with, and potentially buried within, this larger volume of background noise. Consequently, ``burst'' patterns no longer pertain to application-layer behaviors, rendering Wi-Fi flows inherently noisier than TCP/IP flows.

Based on the above insights, we further analyze some existing traffic classification approaches and their limitations in fingerprinting wireless IoT events.

\subsection{Flow-level Analysis}
Flow-based analysis is a promising technique adopted in many network applications, e.g., general traffic analysis~\cite{wan2025cato,Fauvel23LEXNet}, anomaly detection~\cite{jin2009unveiling,zhao2015real,hooshmand2024robust,naseer2018enhanced}, and some network attack detection~\cite{doshi2018machine,idhammad2018semi,nanda2016predicting,churcher2021experimental}. These approaches analyze the statistical/meta data of flows (e.g., average packet sizes, inter-arrival times, flow durations, and etc.), aiming to classify events based on the similarity of repetitive flow-level features. To this end, supervised machine learning (deep learning) is usually adopted for its robustness and accuracy.
While being effective in other applications, applying flow-based ML approaches to Wi-Fi IoT event fingerprinting faces significant limitations.

\begin{figure}[h!] 
\centering
\includegraphics[width=.7\linewidth]{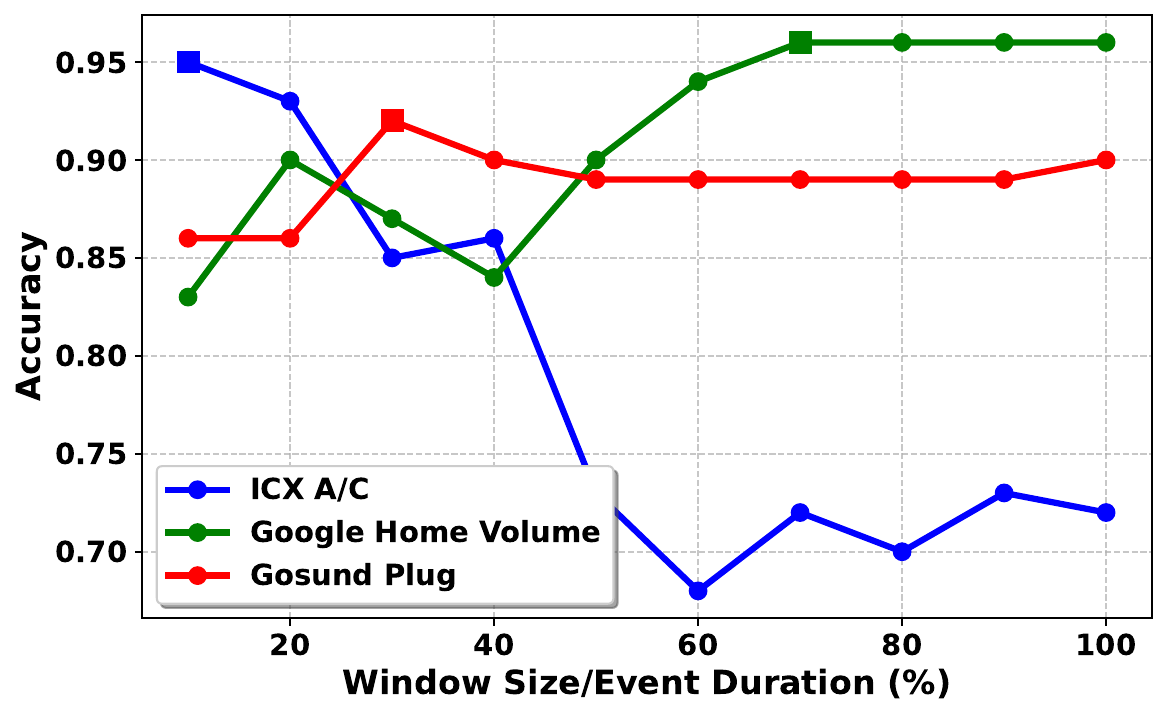}
\caption{\cite{peekaboo} performance with various sliding window and devices/events using chunked training/testing samples.}
\label{fig:peekaboo_cross_device}
\end{figure}

\begin{figure}[h!] 
\centering
\includegraphics[width=.7\linewidth]{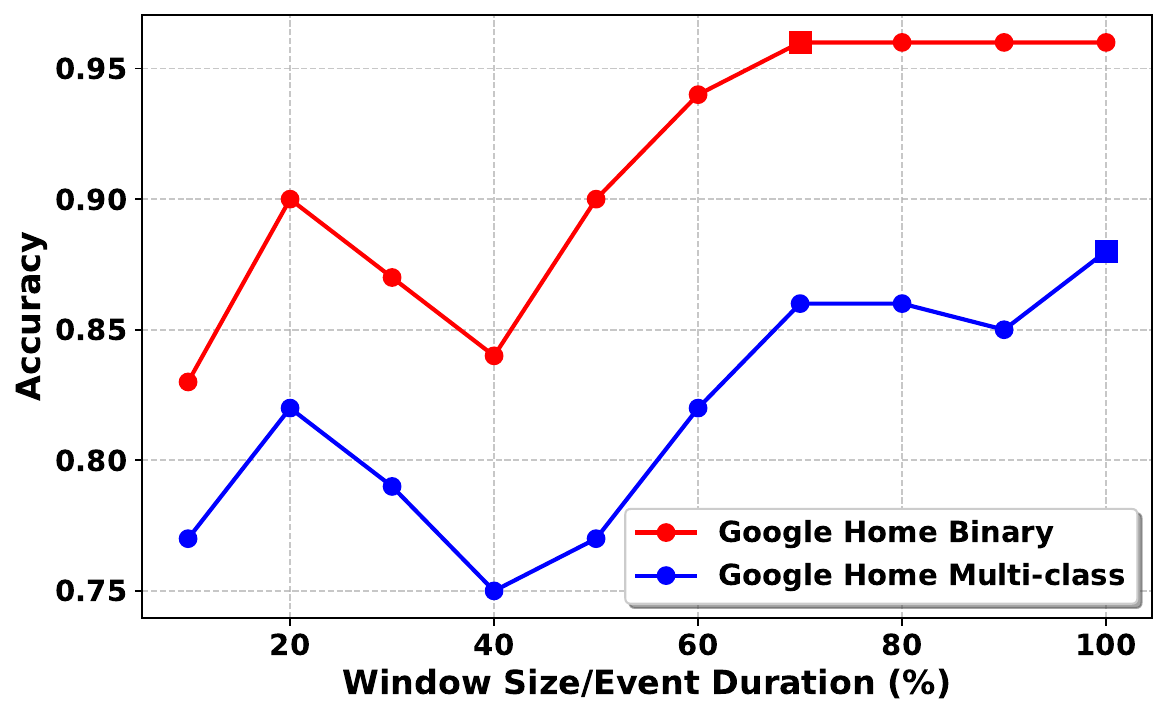}
\caption{\cite{peekaboo} performance of binary classification v.s. multi-event classification using chunked training/testing samples.}
\label{fig:peekaboo_multi_event}
\end{figure}

\smallskip
\noindent \textbf{LIMIT \#1: Collecting labeled training data is too costly considering the countless types of devices in the market place.} The limit of the dataset and the time-consuming data collection process limit the use of deep learning models.
DL models are notable for their capability of extracting high-dimensional features automatically. However, training such models require comparable amounts of labeled data. Existing state-of-the-art DL approaches~\cite{aceto2019mimetic,lotfollahi2020deep,rezaei2019large} mostly focused on classification tasks with large publicly available datasets, which are yet unavailable for Wi-Fi event traffic. 
However, collecting hundreds or thousands of training samples per event is time-consuming (hours to days). Even worse, when a new class (event/device) is introduced, models have to be retrained or finetuned, making the system bulky, considering the rapid development of the IoT domain. 

\smallskip
\noindent \textbf{LIMIT \#2: Parameter tuning and feature set selection have overly significant impact on the overall performance yet low adaptability across devices/events.} For training-efficient machine learning models, expertise for feature selection and hyper-parameter tuning is needed, which are not only labor-intensive, but also suboptimal-prune. Taking the sliding window size as an example, \cite{peekaboo} suggests setting the flow window size as a quarter of the event duration based on empirical observation. Yet, this setting hardly suits every device. We tested the influence of window size on a binary classification task among three devices\footnote{We obtain each device's event duration by roughly calculating the median value of all post-event periods that consist of the majority of packets.} and noticed that the best window size ratio for different devices varies significantly, as shown in Figure~\ref{fig:peekaboo_cross_device}. Considering the enormous amount of features/parameters, testing their combination and tuning parameters for optimal performance is prohibitively inefficient.

\smallskip
\noindent \textbf{LIMIT \#3: Flow-level features are too coarse-grained for multi-event classification.} Some IoT events involve subtle packet size differences (e.g., single-byte variations) too granular for effective capture by aggregate flow-level features~\cite{sivanathan2018classifying,iotsentinel,dong2020lstmfinger,ma2021inferringspatial}. In addition, as mentioned earlier, these subtle differences can be easily obscured by flow-level noise~\cite{homesnitch,ren2019information,copos2016anybody}, such as periodic device heartbeats, or by countermeasures like traffic shaping~\cite{apthorpe2018keepingprivate}. As shown in Figure~\ref{fig:peekaboo_multi_event}, multi-class classification performs much worse than binary classification on two similar events (e.g., Google Home Volume Up \& Down).

\begin{figure}[h!]
\centering
\includegraphics[width=.7\linewidth]{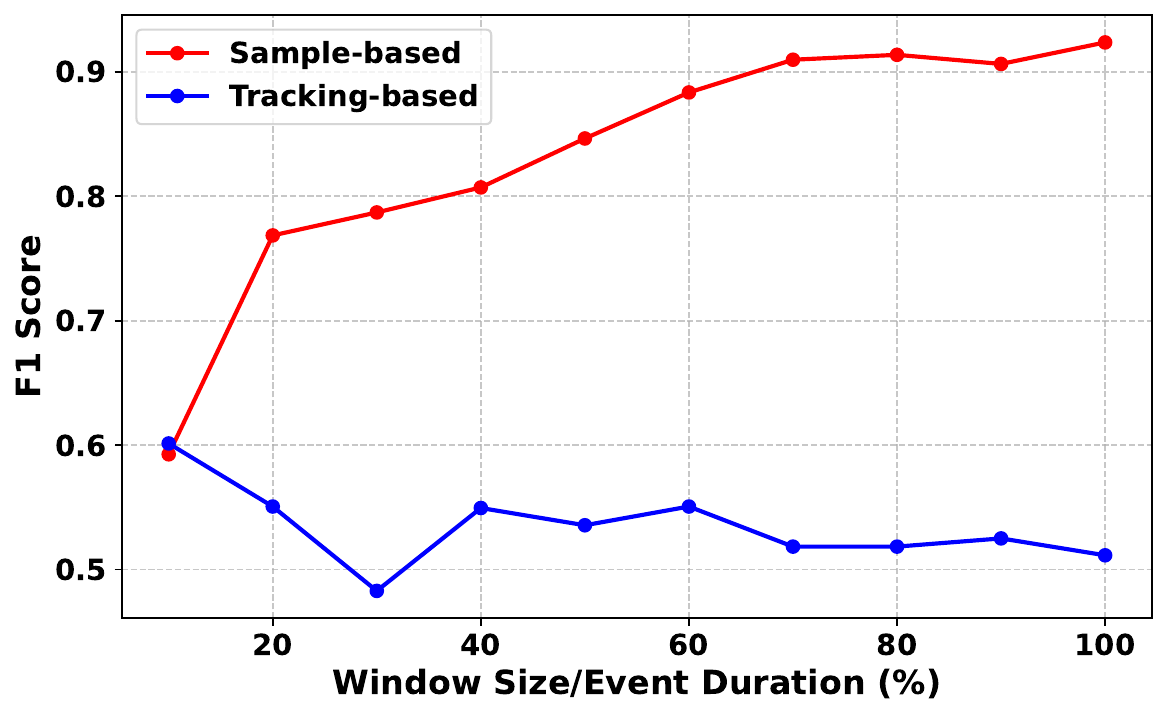}
\caption{F1-Score performances of chunked sample-based detection v.s. continuous tracking. Detailed introductions of ``tracking'' is in Section~\ref{sec:experiment}.}
\label{fig:tracking_comparison}
\end{figure}

\smallskip
\noindent \textbf{LIMIT \#4: ML approaches are evaluated inappropriately.} Most ML methods are overrated due to their inappropriate offline evaluation against chunked samples with accurate labels (isolated flows as samples). However, in a more realistic real-world scenario, classifiers continuously sniffs windows of streaming traffic where they may make decisions with incomplete information. Figure~\ref{fig:tracking_comparison} demonstrates the bias of the two scenarios: classification performance of chunked samples is significantly higher for Google Home events than the more realistic tracking scenario.

\noindent \textbf{LIMIT \# 5: ML approaches for wireless traffic cannot handle the impact of complex events and background noises well.} Some works specifically focus on adapting ML approaches to Wi-Fi traffic~\cite{peekaboo,zou2023iotbeholder,alyami2022wifi}. Nonetheless, without handling Wi-Fi specific noises properly, their performances are degraded upon complex events or noisy background traffic bursts. Meanwhile, some works fingerprint events with other wireless protocol traffic (BLE, Zigbee, Z-Wave~\cite{gu2020iotgaze,zhang2018homonit}). Due to the lower traffic volume and direct data packet-to-application mapping, such protocol traffic is less noisy than Wi-Fi and thus results in better performance. Because of the identical nature of wireless traffic, we only focuses on fingerprinting the more challenging Wi-Fi traffic.

\subsection{Packet-level Analysis}
\label{sec:related:packet-level}
Packet-matching fingerprints events by examining sequences of packet sizes, directions, and time intervals~\cite{wan2021iotathena,pingpong2019trimananda}. Ping-Pong~\cite{pingpong2019trimananda} first utilized unique pairs of packet sizes, stemming from device-server request-response patterns, as event identifiers. Extending this, IoTAthena~\cite{wan2021iotathena} incorporated timing information to form unique packet size sequences with timestamps. Both approaches demonstrated promising results in event fingerprinting and the uniqueness of patterns, surpassing earlier ML-based flow analysis. Unfortunately, both of them cannot operate reliably in Wi-Fi environments.

\begin{figure}[h!] 
\centering
\includegraphics[width=.8\linewidth]{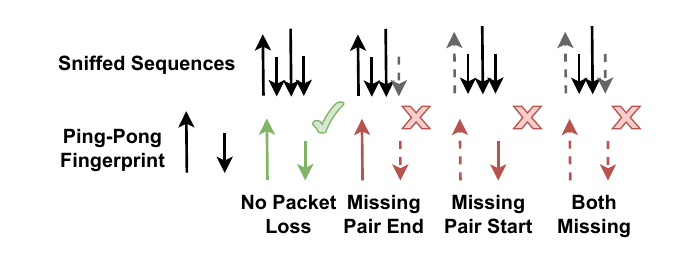}
\caption{Packet-level matching approaches cannot handle packet losses during sniffing.}
\label{fig:match_comparison}
\end{figure}

\smallskip
\noindent \textbf{LIMIT \#6: Packet-level matching approaches are vulnerable to any packet loss and the effects of upper-layer variances.} 
The incompleteness of sniffed traffic significantly impacts online detection methods like \cite{pingpong2019trimananda,wan2021iotathena} due to their exact matching requirements on every single packet. 
Figure~\ref{fig:match_comparison} illustrates how moderate simulated packet loss severely degrades detection performance for \cite{pingpong2019trimananda} and \cite{wan2021iotathena}. 
Furthermore, variances introduced by upper-layer protocols, combined with Wi-Fi layer properties, can interfere with their fingerprint extraction processes. Prior to extraction, Ping-Pong and IoTAthena filter for application-layer data and rely on the pairwise uniqueness of packet sizes to identify those related to specific events.
However, WPA-encrypted Wi-Fi traffic prevents Ping-Pong and IoTAthena from identifying such ``real'' application data.
In large-scale evaluations, we applied Ping-Pong~\cite{pingpong2019trimananda} and IoTAthena~\cite{wan2021iotathena} to Wi-Fi traffic, but both methods exhibit significant scalability challenges without substantial changes and fail to even extract fingerprints.

\subsection{Verbose Information Analysis}
To avoid cumbersome ML feature engineering, some classification approaches~\cite{nprint,saidi2020haystack,apthorpe2017spying,perdisci2020iotfinder} use ``verbose'' raw packet information like headers or packet types (e.g., DNS, ICMP), offering two advantages: (i) reduced feature crafting effort, and (ii) greater insight into upper-layer communication protocols and interactions. For example, to avoid repetitive feature engineering, \cite{nprint} directly encodes raw pcap traffic into bit sequences and deploy Autogluon~\cite{autogluon} as a novel ML pipeline, significantly reducing the effort for data preprocessing. \cite{apthorpe2017spying} identified that some devices always send DNS queries to specific manufactures, easily exposing their brands and types.
However, for Wi-Fi layer IoT event fingerprinting, verbose upper-layer information is generally infeasible as WPA encryption renders payloads indistinguishable.
Furthermore, verbose Wi-Fi frame headers are uninformative for identifying application-layer IoT events. They only reflect local network conditions rather than application behavior.

\section{Problem Statement}
\label{sec:problem}
\subsection{Security \& Privacy Impact}
\label{sec:security_impact}
In this work, attackers seek to accurately track the states of devices of interests by sniffing and analyzing their wireless traffic. Its security and privacy impact is two fold: (i) Curious ``voyeurs'' could use WiFinger to violate individuals' privacy by uncovering reoccurring events that reflect users' living habits or home automation~\cite{house_voyeur,discharge-park-voyeur} such as the time of going to bed or leaving home for work; (ii) Cyber-physical attackers use WiFinger to establish the foundation for the following severer security/privacy violation behaviors, e.g., breaking into \textit{unmonitored} spaces through an \textit{unlocked door}~\cite{burglarasleep,homeinvasion,burglarcamera}. Both situations cause severe privacy and security violations. 

\subsection{Threat Model}
\label{sec:threat_model}
Consistent with prior research~\cite{pingpong2019trimananda,peekaboo,apthorpe2017spying,copos2016anybody,apthorpe2018keepingprivate}, attackers aim to infer IoT device states to deduce users' living habits or gather critical information for subsequent physical intrusions. To achieve this, attackers deploy wireless sniffers in promiscuous mode to capture encrypted traffic. The sniffer could be a compromised IoT device or a self-deployed adapter outside the living space. In either scenario, attackers do not have access to the encrypted payload contents. Nonetheless, since Wi-Fi MAC addresses are in plaintext, attackers can analyze traffic of different devices separately. Before deploying the attack, we assume attackers own some target devices and can collect traffic samples of the target events for training their models. During the attack, attackers need to identify target devices and detect events by analyzing the patterns of anonymous traffic.

To establish more realistic attacking scenarios, attackers identify events from streaming traffic instead of chunked samples, deviating from existing works~\cite{zou2023iotbeholder,peekaboo}. We classify attacks into three categories: naive (binary classification), single-target (targeted event detection), and multi-target (multi-event monitoring). In the naive scenario, attackers can merely separate a single target event from the idle state, being useful in limited contexts (e.g., inferring a light is ``off'' as the last event late at night). In the single-target scenario, attackers build fingerprints/models using multi-event training data but aim to distinguish only one specific event from all other events and idle traffic. This is helpful when targeting critical events (e.g., door unlocking) that are important/sensitive and may occur unpredictably. In the most advanced multi-target scenario, attackers aim to monitor and identify multiple target events occurring on a device. This is also the most challenging case, as misclassified events can impact subsequent detection results. Despite its challenges, multi-target tracking provides the most information and is thus the ultimate goal of such privacy inference attacks.

\subsection{Connection Configurations \& Event Triggering}
There are two main types of connection configurations for wireless smart devices: they either connect directly to the home router via Wi-Fi or connect to a smart hub (e.g., Amazon Echo, Google Home) which then connects to the router. Devices with the former configuration usually consume more power and often have more advanced functions. In contrast, the latter is more common for BLE or Zigbee devices that stay idle most of the time. In this work, we focus primarily on the first type of connection due to its prevalence and complexity, and our approach could be easily adapted to the hub-based configuration by sniffing communications between the hub and devices. Moreover, there are two types of event triggering schemes: via pre-configured smart home automation or via manual operations on companion apps. Since the two schemes result in similar traffic patterns (similar conclusions in \cite{pingpong2019trimananda}), we focus on the latter scenario, primarily for data collection efficiency and scalability.

\section{WiFinger: System Design}
\label{sec:wifinger}
In this section, we introduce our packet-level fingerprinting approach, WiFinger. Starting with an analysis of Wi-Fi traffic noise, we demonstrate the intuition behind effective packet-level fingerprinting. In the following subsections, we first define the ideal packet-level fingerprint for noisy traffic and discuss how to efficiently deploy online fingerprint matching. Subsequently, we address the challenges of obtaining such ideal fingerprints.

\subsection{Intuition on Packet-level Fingerprinting}
Ping-Pong~\cite{pingpong2019trimananda} initially proposed the use of unique data packet pair sizes as fingerprints. IoTAthena~\cite{wan2021iotathena} further demonstrated the uniqueness of using packet sizes and intervals as the matching criterion among various IoT devices.
When such data packets are transmitted over Wi-Fi, they are encapsulated within Wi-Fi data frames, adding new headers, footers, and potential padding. Although WPA encryption hides the upper-layer packet structures, observable frame characteristics like size, direction, and timing remain available. We posit that the sequence of Wi-Fi data frames corresponding to an event, despite encapsulation overheads, retains a characteristic pattern of relative timings and frame sizes derived from the original TCP/IP exchange.

\begin{figure}[h!]
    \centering
    \includegraphics[width=.8\linewidth]{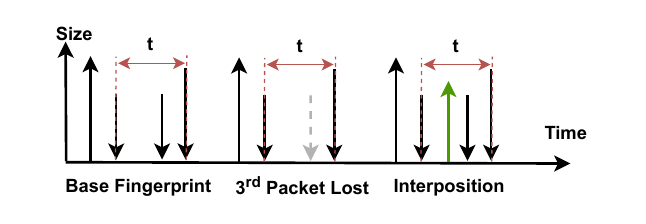}
    \caption{Time intervals between packets remain relatively consistent when there are packet losses or interpositions.}
    \label{fig:consis_inter}
\end{figure}

Consequently, in an ideal scenario with no packet loss at either layer, a Wi-Fi IoT event can be characterized by a sequence of Wi-Fi data packets defined by their sizes, directions, and time intervals. We term this ideal Wi-Fi traffic sequence the \textbf{\textit{base fingerprint}} and the constituent packets as \textbf{\textit{fingerprint packets}}. Upon an event occurrence, two key deviations from base fingerprints are observed:
First, occasional packet loss during an event can result in the removal of \textit{fingerprint packets}; second, the encapsulation of irrelevant packets or variations in the network environment can introduce additional packets (noise) into the observed traffic stream alongside the base fingerprint packets.
Regardless of the noise, some or all \textit{fingerprint packets} will remain within the event traffic, maintaining relatively stable inter-arrival times, as illustrated in Figure~\ref{fig:consis_inter}. Therefore, the event traffic classification problem can be reformulated as determining whether a given time series sequence partially matches the \textit{base fingerprint}. In what follows, we discuss our resolutions to two challenges: \textbf{1. Matching base fingerprints efficiently; 2. Extracting base fingerprints from incomplete traffic}.

\subsection{Fingerprint Matching}
\label{subsec:fuzzy_match}
First, assuming that base fingerprints have been obtained, we conceptualize the traffic matching problem as a variation of the Longest Common Subsequence (LCS) problem, predicated on packet transmission directions, sizes, and interval constraints.  The closer the match between two sequences based on this criterion, the more likely the target sequence corresponds to the same event. We define the \textbf{l}ongest \textbf{c}ommon \textbf{s}ubsequence of \textbf{n}etwork \textbf{t}raffic (NT-LCS) problem as follows:

\begin{definition}[NT-LCS]
\label{def:fmlcs}
For input sequences $Seq_1$ (base fingerprint) and $Seq_2$ (sequence to be examined), the NT-LCS is the longest subsequence between the two sequences:
\begin{equation*}
\begin{aligned}[t]
    & \Bigl[ p^a_1, p^a_2, \dots, p^a_n \Bigr], \quad p^a_i \in Seq_1^{sub}, \quad Seq_1^{sub} \subseteq Seq_1 \\
    & \Bigl[ p^b_1, p^b_2, \dots, p^b_n \Bigr], \quad p^b_i \in Seq_2^{sub}, \quad Seq_2^{sub} \subseteq Seq_2
\end{aligned}
\end{equation*}
where each element $p$ contains:
\begin{align*}
    p^a_i &= \{\,\text{time}:a_i,\ \text{size}:s^a_i,\ \text{dir}:d^a_i\,\} \\
    p^b_i &= \{\,\text{time}:b_i,\ \text{size}:s^b_i,\ \text{dir}:d^b_i\,\}
\end{align*}
subject to:
\begin{itemize}
    \item $\sqrt{\sum_{i=1}^n (a_i-b_i)^2} \leq \beta$ (time constraints)
    \smallskip
    \item $(s_i^a-s_i^b)^2 \leq \epsilon^2 \quad \forall i \in [1,n]$ (similar packet size)
    \smallskip
    \item $d_i^a=d_i^b \quad \forall i \in [1,n]$ (same direction)
\end{itemize}
\end{definition}

Since the NT-LCS problem is NP-hard (proof in Appendix~\ref{sec:app:np_hard}), we propose a baseline approximation algorithm, FMLCS (\textbf{F}uzzily \textbf{M}atching for \textbf{L}ongest \textbf{C}ommon \textbf{S}ubsequence). FMLCS extends the dynamic programming (DP) approach for LCS, adapted to accommodate permissible variations in packet size and time intervals, as detailed in Algorithm~\ref{algo:fmlcs}. In the DP table construction, we use one table ($max_{tab}$) to track common subsequence lengths and another ($lcss_{tab}$) to record possible subsequences corresponding to the lengths at each step. During the DP function, two packets are matched if they share the same transmission direction and their size difference is less than or equal to $\epsilon$ bytes (line 30). After identifying potential longest common subsequences (lines 6-9), their temporal alignments with the \textit{base fingerprint} are measured by calculating the L2Norm distances between the relative timestamps of the matched packets (lines 17-24). The subsequence with the best temporal alignment (minimum distance) is selected as the matching result (lines 10-12). For a match to be deemed successful, the selected subsequence's length must be at least $\gamma$\% of the base fingerprint's length and exhibit strong temporal alignment ($dist \leq \beta$). During experiments, we set $\epsilon=1$, $\gamma=0.6$ (60\% of the base fingerprint), and $\beta=2$. These values serve as the thresholds for successful matching (line 13).

\begin{algorithm}[h]
    \small
    \caption{Baseline FMLCS}
    \label{algo:fmlcs}
    \begin{algorithmic}[1]
        \algrenewcommand{\algorithmicindent}{8pt} 
        \algtext*{EndIf}
        \algtext*{EndFor}
        \algtext*{EndFunction}
        
        \State \textbf{Input:} $Seq_1$ (base fingerprint), $Seq_2$ (target sequence)
        \State \textbf{Output:} $true$/$false$ (whether $Seq_2$ matches $Seq_1$)
        \Statex
        
        \Function{FMLCS}{$Seq_1$, $Seq_2$}
        \State $max_{len}, lcss_{tab}$ = $DP(Seq_1, Seq_2)$ 
        \State $dist_{min}$ = $\infty$, $Seq_{match}$ = $[]$
        \For {$lcs \in lcss_{tab}$}
        \If{$len(lcs) < max_{len}$}
        \State \textbf{continue}
        \EndIf
        \State $dist_{temporal}$ = $TimeAlignment(lcs)$
        \If{$dist_{temporal} < dist_{min}$}
        \State $dist_{min}$ = $dist_{temporal}$
        \State $Seq_{match}$ = $lcs$
        \EndIf
        \EndFor
        \If{$len(Seq_{match})\geq \gamma*len(Seq_1)$ \& $dist_{min} < \beta$}
        \State \textbf{return} $true$
        \Else
        \State \textbf{return} $false$
        \EndIf
        \EndFunction
        
        \Function{TimeAlignment}{$lcs$}
        \State $Subseq_1$ = $Seq_1$[$index[1]$ for $index$ in $lcs$]
        \State $Subseq_2$ = $Seq_2$[$index[2]$ for $index$ in $lcs$]
        \State $time\_vec_1$ = [$packet[time]$ for $packet$ in $Subseq_1$]
        \State $time\_vec_2$ = [$packet[time]$ for $packet$ in $Subseq_2$]
        \State $time_1$ = $time\_vec_1 - mean(time\_vec_1)$
        \State $time_2$ = $time\_vec_2 - mean(time\_vec_2)$
        \State \textbf{return} $l2norm(time_1 - time_2)$
        \EndFunction
        
        \Function{DP}{$seq_1, seq_2$}
        \State $paths$ = Empty list
        \State $max_{tab} = [[0]*len(seq_1)]*len(seq_2)$
        \State $lcss_{tab} = [[paths]*len(seq_1)]*len(seq_2)$
        \For {$i$=1 to $len(seq_1)$}
        \For {$j$=1 to $len(seq_2)$}
        \If{$seq_1[i] \approx seq_2[j]$} \textcolor{blue}{\# match ``common'' packets}
        \State $max_{tab}[i][j] = max_{tab}[i-1][j-1]+1$
        \State $path_{all} = lcss_{tab}[i-1][j-1]$
        \For{$path$ in $path_{all}$}
        \State $path_{all}.append([(i,j)])$
        \EndFor
        \State $lcss_{tab}[i][j]$ = $path_{all}$
        \Else
        \If{$max_{tab}[i-1][j] > max_{tab}[i][j-1]$}
        \State $max_{tab}[i][j]$ = $max_{tab}[i-1][j]$
        \State $lcss_{tab}[i][j]$ = $lcss_{tab}[i-1][j]$
        \ElsIf{$max_{tab}[i-1][j] < max_{tab}[i][j-1]$}
        \State $max_{tab}[i][j]$ = $max_{tab}[i][j-1]$
        \State $lcss_{tab}[i][j]$ = $lcss_{tab}[i][j-1]$
        \Else
        \State $max_{tab}[i][j]$ = $max_{tab}[i-1][j]$
        \State $lcss_{tab}[i][j]$ = $lcss_{tab}[i][j-1] + lcss_{tab}[i][j-1]$
        \EndIf
        \EndIf
        \EndFor
        \EndFor
        \State \textbf{return} $max_{tab}[-1][-1], lcss_{tab}$
        \EndFunction
    \end{algorithmic}
\end{algorithm}

\begin{figure}[h!]
    \centering
    \includegraphics[width=.8\linewidth]{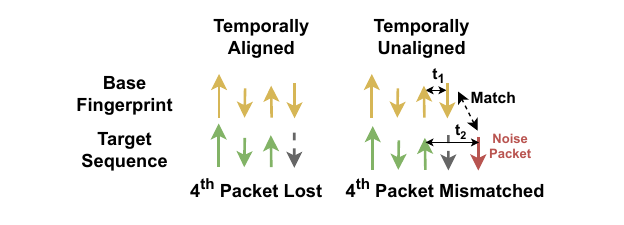}
    \caption{Erroneous packet-matching predicated only on sizes and directions, leading to temporal unalignment.}
    \label{fig:acc_mismatch}
\end{figure}

For smart devices with simple commands (e.g., plugs, lights, A/C controllers), FMLCS is already sufficient to detect IoT events accurately and efficiently. However, for complex devices with larger volumes of traffic (e.g., Amazon Echo Dot), this baseline has two severe limitations.
\textbf{Computation efficiency:}
The algorithm is computationally expensive as the time complexity of DP increases with the target sequence length. Furthermore, obtaining all possible longest common subsequences requires either recording all matches during DP (current implementation) or recursive backtracking (which is even slower); both consume unacceptable time for long target sequences.
\textbf{Accidental mismatch:}
In rare cases, IoT events can be mismatched. This occurs because FMLCS only checks temporal alignment for the longest common subsequences. If a \textit{fingerprint packet} is not event-unique, accidentally matched noise packets can extend the maximum length of common subsequences and consequently impact the final temporal alignment, as shown in Figure~\ref{fig:acc_mismatch}. To tackle these problems, we further analyze the characteristics of complex event fingerprints and propose a more advanced approximation algorithm.

\subsection{Advanced Fingerprint Matching}
\label{sec:blocked_fmlcs}
The challenge of computation costs and fingerprint mismatches stems from the same root cause: the current common packet-matching step (line 30) considers only sizes and directions, not time. Consequently, if many packets share similar sizes and directions but different timestamps, the number of subsequence combinations increases exponentially, even resulting in the erroneous matching of temporally unaligned packets. To address this issue, we propose two key optimizations to fully leverage temporal information: Anchor Reference and Fingerprint Segmentation.

\begin{figure}[h!]
    \centering
    \includegraphics[width=.8\linewidth]{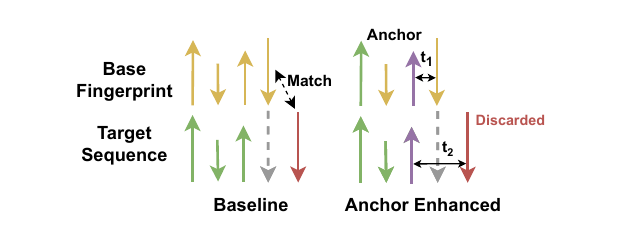}
    \caption{Using anchor packets to filter out erroneous packet-matching, i.e., $abs(t_1-t_2)>\alpha$.}
    \label{fig:anchor_packet}
\end{figure}

\textbf{Anchor Reference:}
Given an IoT event, intervals between fingerprint packets remain relatively steady. This interval steadiness can be utilized to filter unnecessary packet-matching during DP, as shown in Figure~\ref{fig:anchor_packet}. Specifically, we start by stochastically choosing one of the fingerprint packets as the anchor and try to match it using size and direction. Once discovered, these two packets serve as anchor packets for all subsequent potential matches by constraining their time intervals relative to the anchor packets, i.e., $abs(t_1-t_2)\leq \alpha$. Considering network fluctuation, empirically setting $\alpha$ to 0.2s is sufficient to accommodate regular network jitters. Nonetheless, this leaves opportunities for knowledgeable defenders to circumvent WiFinger by deliberately delaying packets to increase the intervals beyond the threshold. To alleviate this issue, we further leverage the interval distribution of packet pairs to dynamically calibrate packet intervals before anchor-based matching. Specifically, we first record the stable distribution of intervals of all consecutive packets (denoted as pairs) without interleaving missing packets. Then, given a newly sniffed sequence, WiFinger rectifies the pair-level intervals to ensure they are within 1-sigma deviations from the stable distribution. In addition, packet intervals larger than 0.5s are excluded from adjustments because such substantial intervals either indicate natural burst separations or severe functioning degradation due to defenses. In Section~\ref{sec:countermeasures}, we evaluate WiFinger with and without interval calibration (IC) against the deliberate delaying defense separately.

\textbf{Fingerprint Segmentation:}
For high traffic-volume devices, we additionally introduce \textit{segmented matching} based on a critical observation: packets within long base fingerprints further exhibit temporal clustering patterns at finer time scales. This is because complex IoT events typically involve multi-round communications, manifesting as several small packet bursts (segments). Therefore, we decompose the full-sequence matching task into parallelizable subtasks, each targeting one segment of the \textit{base fingerprint}. To achieve this, we divide the base fingerprint using consecutive packet intervals, with boundaries set at the middle of intervals exceeding a threshold of $0.5$s. Segments containing fewer than three packets are merged into subsequent segments. Finally, we apply the anchor-constrained FMLCS to all segments independently and aggregate the matching results (Figure~\ref{fig:seg_match}). This approach effectively mitigates the combinatorial explosion in DP-based packet matching that scales exponentially with sequence length. 
Overall, the advanced FMLCS (AFMLCS) works as follows:
\begin{figure}[h!]
    \centering
    \includegraphics[width=.8\linewidth]{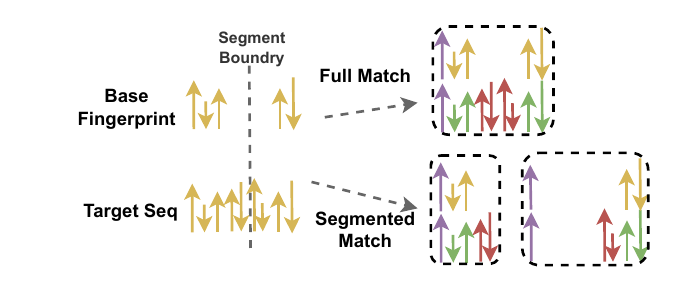}
    \caption{Segmentation reduces the number of processed packet in each segment, decreasing the computation costs.}
    \label{fig:seg_match}
\end{figure}
\begin{itemize}[leftmargin=*]
\item \textbf{1. Fingerprint segmentation:} A long base fingerprint is divided into several segments.
\item \textbf{2. Anchor packet selection:} WiFinger sequentially compares fingerprint packets within the first segment with the target sequence to identify the first matching packet pair as anchors. If no such match is found, the remaining matching process is terminated.
\item \textbf{3. Match the first segment:} The anchor-constrained FMLCS is applied to the first segment and the target sequence. If this matching fails, matching for the remaining segments is terminated.
\item \textbf{4. Merge segmented matching results:} After a successful match for the first segment, the same anchor-constrained matching is applied to all remaining segments, whose results will get merged. Such a final result is successful if it involves at least $\gamma$\% of the base fingerprint packets and they align well temporally (e.g., meeting the $\beta$ distance criterion).
\end{itemize}

\subsection{Fingerprint Extraction}
\label{sec:fp_extraction}
\begin{figure}[h]
    \centering
    \includegraphics[width=\linewidth]{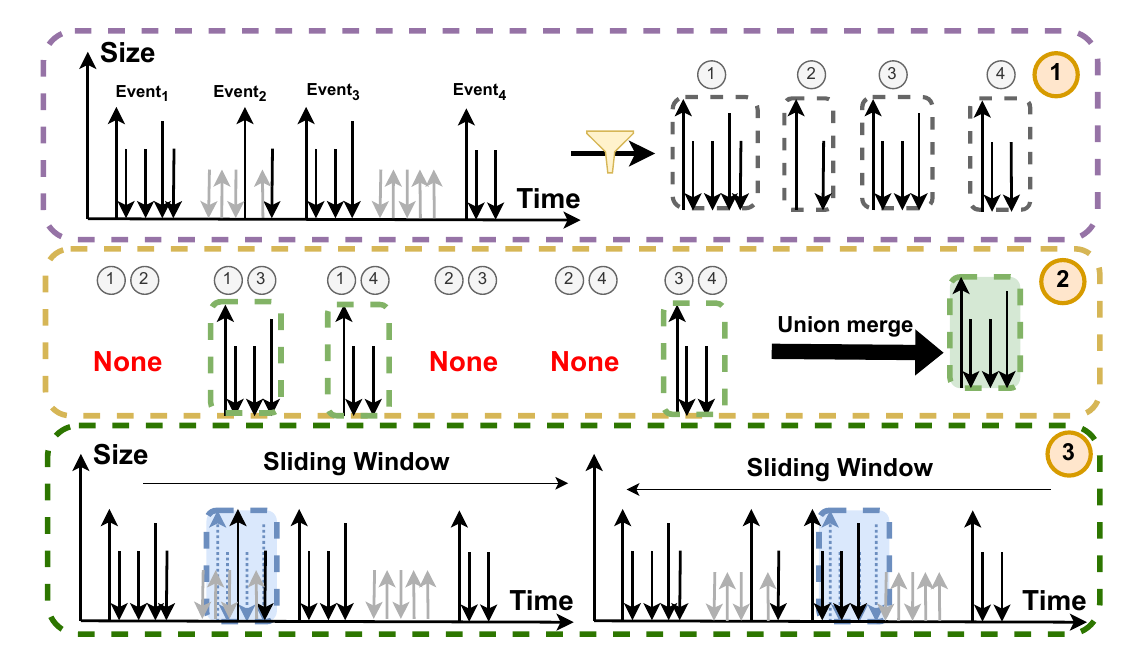}
    \caption{Fingerprint extraction process. Step 1 filters out noise packets and clusters packets into groups. Step 2 applies pairwise AFMLCS to the groups and merges their consensus subsequences. Step 3 refines the coarse fingerprints by measuring packets' matching frequencies against the training data.}
    \label{fig:fp_extraction}
\end{figure}

While the matching approach relies on accurate \textit{base fingerprints} for detection, \textbf{acquiring clean and reliable ones presents fundamental challenges}: ideal base fingerprints must be derived from noisy training data exhibiting high packet losses (5-20\% observed rates) and upper-layer variations. To address this, our solution exploits the \textit{collective intelligence} of repeated event executions: each noisy event traffic burst is assumed to partially manifest the base fingerprint sequence.  Therefore, we formulate base fingerprint extraction as another subsequence matching problem across all recorded event traffic bursts. Unlike the online matching paradigm, subsequence matching is used differently during extraction. We first use FMLCS to extract pairwise common subsequences among all ``corrupted'' traffic bursts to obtain a set of potential fingerprint components. Then, we merge these components into one noisy coarse fingerprint (CF) and leverage statistics of packet matching frequencies to refine the CF into the base fingerprint. Overall, the fingerprint extraction process consists of three steps: (i) data collection and filtering; (ii) coarse fingerprint construction; and (iii) fingerprint refinement, as illustrated in Figure~\ref{fig:fp_extraction}.

\subsubsection{Data Collection, Filtering, and Compression}
\label{sec:data_filter_compress}
To obtain the training dataset, we use ADB (Android Debug Bridge)~\cite{adb} and Python scripts to build an automatic event execution tool, simulating taps or swipes on mobile devices~\cite{pingpong2019trimananda}. The tool initiates a specific event $X$ times (30 in experiments) to trigger traffic generation from devices/servers. During automated data collection, the tool records timestamps of event initiations (taps/slides) and captures the Wi-Fi packets exchanged between the device and the router. Random gaps of 30-45 seconds are introduced between each event to ensure the device finishes the event and returns to the idle state.

After data collection, we first discard Wi-Fi Management and Control packets, focusing only on Wi-Fi Data packets, as discussed in Section~\ref{sec:example_fingerprint}. To avoid collecting repeated data packets of Wi-Fi retransmissions, only the original packet or one of its retransmission packets is kept. For devices with large traffic volumes, we further filter potential irrelevant noise packets to reduce computation costs. First, packets are categorized into classes by their sizes and directions, and those classes whose frequencies are significantly higher than the average frequency (e.g., exceeding it by 2$\sigma$ or more) are discarded. These exceptionally frequent classes likely correspond to background noise, such as TCP ACK packets or regular data uploads. Then, we compress consecutive packets with the same sizes and directions into one packet. This is because identically sized packets lead to the combinatorial explosion challenge in FMLCS, but they do not contribute to the characteristics of request-response patterns of IoT events~\cite{pingpong2019trimananda}. Such a compression significantly reduces the number of packets to be processed, e.g., from 10000 to 3000 for Amazon Echo Dot. Finally, according to~\cite{pingpong2019trimananda}, IoT events typically last no more than 10 seconds. Therefore, packets sent beyond a 15-second window following event initiations are discarded. In the end, we obtain $X$ groups of 15-second packet traces, each corresponding to an event.

\subsubsection{Coarse Fingerprint Extraction}
We use the $X$ groups of preprocessed traffic to construct a coarse fingerprint (CF). A CF is expected to contain the base fingerprint but may also include some noise packets. To extract a CF, we leverage the \textit{collective intelligence} of traffic groups: we extract consensus subsequences from every pair of traffic groups and take the union of these consensus subsequences as the CF. To this end, we further adapt AFMLCS to reduce the computation costs of the extraction process.

\noindent \textbf{Extraction Adaptations:} The key difference between the matching and extraction phases is the noisiness of the two sequences being compared. In the matching phase, the base fingerprint is assumed to be noise-free. Thus, the naive anchor packet selection (based only on packet sizes and directions) does not severely sacrifice efficiency or accuracy. However, when both sequences are noisy, selected anchors for both sequences are very likely to be mismatches, leading to erroneous references for all subsequent packets. Even worse, during extraction, such mismatched subsequences (containing significant noise) will also be treated as potential fingerprint components, consequently resulting in a significant increase in computational cost and considerable noise in the final CF.

\begin{figure}[h!]
    \centering
    \includegraphics[width=.8\linewidth]{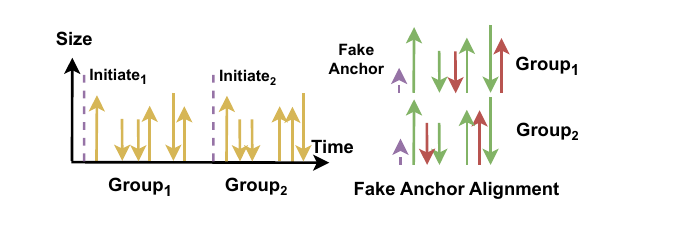}
    \caption{Align traffic groups by their event-initiation timestamps (fake anchor packets) during the extraction.}
    \label{fig:fake_anchor}
\end{figure}

To address this, we insert fake packets at the event initiation timestamps to serve as anchor packets. Due to stable network connection of the attackers' testbed, traffic bursts emerge with a relatively fixed delay after event initiations. Since the intervals between fingerprint packets within the burst are also steady, event initiation timestamps serve effectively as reference anchors in AFMLCS. As such, fake identical packets (anchors) are inserted at each event initiation timestamp to help align sequences temporally, as shown in Figure~\ref{fig:fake_anchor}. After obtaining all pairwise common subsequences, we naively merge all subsequences into a coarse fingerprint.

\subsubsection{Fingerprint Refinement}
Due to the rudimentary union merging, the CF inadvertently incorporates redundant noise packets. To filter such noise, we refine the fingerprint by conducting AFMLCS against all groups using a sliding window scheme, while meticulously recording the matching frequencies for the packets inside the CF. This aims to effectively pinpoint CF packets whose matching frequencies approximately align with the number of events, i.e., $X$. While a unidirectional sliding window might result in elevated matching frequencies for packets located at the beginning or end of the sequence, we execute the sliding window analysis in both directions. In our bidirectional refinement, packets with matching frequencies less than $X$ are discarded. We iteratively execute the refinement process until no further packets are discarded, and construct the base fingerprint with all retained packets.

\subsection{System Workflow}
\label{sec:workflow}
\begin{figure}[h!]
    \centering
    \includegraphics[width=.8\linewidth]{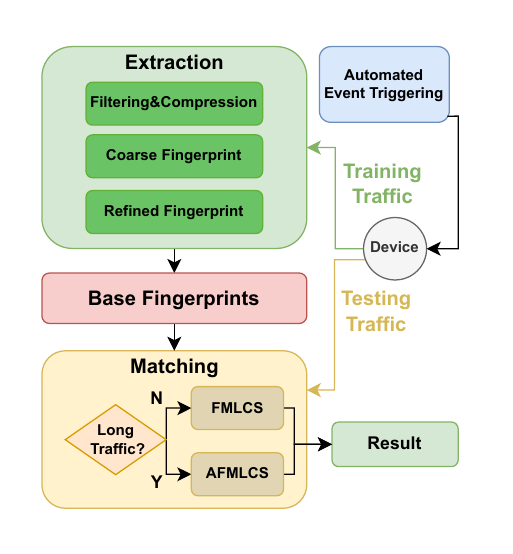}
    \caption{WiFinger system workflow.}
    \label{fig:wifinger}
\end{figure}

In general, the workflow for WiFinger is shown in Figure~\ref{fig:wifinger}. During the extraction phase, WiFinger uses the event triggering module to trigger each event 30 times, collects the corresponding Wi-Fi traffic, and extracts base fingerprints for the events. In the online detection phase, WiFinger selects the target base fingerprint and matches it to newly sniffed device traffic using either FMLCS or AFMLCS, depending on the traffic volume and the fingerprint length.

\section{Evaluation}
\label{sec:experiment}
We conduct experiments corresponding to the three attacking scenarios (Section~\ref{sec:problem}) to evaluate WiFinger's performance against two state-of-the-art ML-based fingerprinting methods applied to Wi-Fi traffic: Peek-a-boo~\cite{peekaboo} and IoTBeholder~\cite{zou2023iotbeholder}. For Peek-a-boo, we use the same feature set and select their best-performing model (Random Forest, or RF) for evaluation. During the experiments, we assume RF and IoTBeholder have finished their device classification and only focus on the event detection. As for WiFinger, we aim to use a single fingerprint to classify devices and events at the same time. \footnote{Ping-Pong and IoTAthena are not included as baselines due to their feasibility on Wi-Fi traffic, as discussed in Section~\ref{sec:related:packet-level}}

\subsection{Dynamic Tracking \& Evaluation Metrics}
\label{sec:detection_metrics}
We use continuous event tracking to emulate realistic attacks instead of classifying chunked traffic flow samples. Specifically, every detection method uses a sliding window to dynamically select a group of most recent packets for classification.  For Peek-a-boo, we test various window sizes and select the best-performing setting for each event. For IoTBeholder, the window size is set as the burst duration, using the same definition as~\cite{zou2023iotbeholder}. For WiFinger, the window size is set as the duration of the extracted base fingerprint plus two seconds (to accommodate potential timing variations). Given a window of packets, each method determines whether it corresponds to ``idle'' (negative) or an event (positive). Whenever an event is detected, packets within the current window are excluded from subsequent windows to avoid misclassification of similar events. Specifically, Peek-a-boo and IoTBeholder skip all packets in the next 6 seconds, while WiFinger skips all packets contained within the current window. This setting is practical as IoT events on the same device seldom occur consecutively within a very short period, and the collected events in our dataset have long enough gaps in between.

To evaluate dynamic tracking performance, we use precision and recall rates as metrics. During detection, true positive results correspond to events with correct labels, and false positive results correspond to events with wrong labels or misclassifications of idle states. Precision is defined as the ratio of true positive detections to the total number of positive detections. Recall is defined as the ratio of true positive detections to the total number of triggered events.

\subsection{Attacking Scenarios}
\label{sec:scenarios} 
We evaluate methods' dynamic tracking performance in three scenarios: naive, single-target, and multi-target tracking (ultimate objective).\footnote{ It is worth noting that some event fingerprints (E11-E13, E18-19) are inseparable; they are excluded from the single-target/multi-target experiments.} For Peek-a-boo and IoTBeholder, attackers use various strategies to train models and detect events to achieve the optimal performance.
\begin{itemize}[leftmargin=*]
 \item \textbf{Naive:} attackers train simple binary classifiers to distinguish one target event from idle periods.
 \item \textbf{Single-target:} attackers train multi-event classifiers for each device and aim to distinguish only the event of interest from all other events and idle, i.e., attackers discard any report of non-target events.
 \item \textbf{Multi-target:} attackers train multi-event classifiers to monitor all occurring events of a device, i.e., attackers accept reports of any event.
\end{itemize}
As for WiFinger, attackers use a consistent procedure to extract event fingerprints and apply (A)FMLCS to match them in all three scenarios. 

\subsection{Testbed Configurations \& Dataset Collection}
\label{sec:testbed} 
We built an automated event triggering system using a XiaoMi 8 mobile device to trigger IoT events. Traffic generated by the devices was sniffed and labeled accordingly.  We tested the sniffing performance of a MacBook, a NetGear A6210 adapter~\cite{netgear}, and an ALFA AWUS036ACH adapter~\cite{rtl8812au}, and finally chose the A6210 running on Ubuntu-16.04 for its best capture performance (lowest packet loss). For each IoT event, we generated 30 samples in a lab environment for training and 20 samples in a home environment for testing. All samples had approximately 40-second gaps between them to ensure completion of events. In total, our dataset includes 15 devices and 47 events (Table~\ref{tab:events}) representative of smart devices on the market, including smart home agents, small smart peripherals, and integrated smart actuators. Devices with simpler functionality are TP-Link Plug, Gosund Plug, WAH Plug, ICX-RF Controller, and Wiz Hue Light. For complex devices, we chose Amazon Echo, Google Home, Xiaomi Smart Sweeper and humidifier, Midea Dishwasher and Dish Sterilizer, Sprinkler, two thermostats, and a ring alarm. Apart from the tested devices, a laptop and a TV were connected to the same network serving as background noise traffic. All devices were connected via 2.4GHz Wi-Fi.

\subsection{End-to-End Detection Results}
The results of the naive binary classification are shown in Table~\ref{tab:naive}. Under the naive setting, Peek-a-boo obtains the highest recall rate of 96\% and WiFinger achieves the highest precision rate of 98\% on average.
\begin{table}[h]
\centering
\begin{tabular}{|c|cc|cc|cc|}
\hline
\multirow{2}{*}{Methods} & \multicolumn{2}{c|}{Peek-a-boo RF}& \multicolumn{2}{c|}{IoTBeholder}& \multicolumn{2}{c|}{WiFinger(ours)}\\ \cline{2-7} & \multicolumn{1}{c|}{Rec.}& Prec.& \multicolumn{1}{c|}{Rec.}& Prec.& \multicolumn{1}{c|}{Rec.}& Prec.\\ \hline
Average& \multicolumn{1}{c|}{\textbf{0.95}} & 0.88 & \multicolumn{1}{c|}{0.9} & 0.9 & \multicolumn{1}{c|}{0.90} & \textbf{0.98} \\ \hline
\end{tabular}
\caption{Three models demonstrate similar performance on the naive scenario.}
\label{tab:naive}
\end{table}

Advancing to the single-target setting, WiFinger maintains excellent performance, but Peek-a-boo and IoTBeholder demonstrated a decreasing trend. As shown in Table~\ref{tab:regular}, Peek-a-boo achieves 83\% recall and 81\% precision, while IoTBeholder achieves 74\% recall and 87\% precision. The performance gap between both models and WiFinger mainly lies in the precision. The noisiness of WiFi traffic increases the difficulty of distinguishing events especially when their fingerprints are similar. For example, E7 and E8 (also E9 and E10) only have byte-level differences inside a window, causing obvious degradation on Peek-a-boo and IoTBeholder.
\begin{table*}[h]
 \centering
 \footnotesize
 \renewcommand{\arraystretch}{1.1} 

\begin{tabular}{|c|cccc|cccc|cc|}
\hline
\multirow{3}{*}{Event ID} &
 \multicolumn{4}{c|}{Single-target} &
 \multicolumn{4}{c|}{Multi-target} &
 \multicolumn{2}{c|}{Single-target/Multi-target} \\ \cline{2-11} 
&
 \multicolumn{2}{c|}{Peek-a-boo RF} &
 \multicolumn{2}{c|}{IoTBeholder} &
 \multicolumn{2}{c|}{Peek-a-boo RF} &
 \multicolumn{2}{c|}{IoTBeholder} &
 \multicolumn{2}{c|}{WiFinger(ours)} \\ \cline{2-11} 
&
 \multicolumn{1}{c|}{Recall} &
 \multicolumn{1}{c|}{Precision} &
 \multicolumn{1}{c|}{Recall} &
 Precision &
 \multicolumn{1}{c|}{Recall} &
 \multicolumn{1}{c|}{Precision} &
 \multicolumn{1}{c|}{Recall} &
 Precision &
 \multicolumn{1}{c|}{Recall} &
 Precision \\ \hline
E1 &
 \multicolumn{1}{c|}{0.9} &
 \multicolumn{1}{c|}{1} &
 \multicolumn{1}{c|}{0.2} &
 0.67 &
 \multicolumn{1}{c|}{0.95} &
 \multicolumn{1}{c|}{1} &
 \multicolumn{1}{c|}{0.15} &
 0.3 &
 \multicolumn{1}{c|}{0.9} &
 0.72 \\ \hline
E2 &
 \multicolumn{1}{c|}{1} &
 \multicolumn{1}{c|}{0.9} &
 \multicolumn{1}{c|}{1} &
 1 &
 \multicolumn{1}{c|}{0.05} &
 \multicolumn{1}{c|}{0.04} &
 \multicolumn{1}{c|}{0.25} &
 0.25 &
 \multicolumn{1}{c|}{1} &
 1 \\ \hline
E3 &
 \multicolumn{1}{c|}{0.95} &
 \multicolumn{1}{c|}{1} &
 \multicolumn{1}{c|}{0.95} &
 1 &
 \multicolumn{1}{c|}{0.15} &
 \multicolumn{1}{c|}{0.11} &
 \multicolumn{1}{c|}{0.1} &
 0.07 &
 \multicolumn{1}{c|}{0.95} &
 1 \\ \hline
E4 &
 \multicolumn{1}{c|}{1} &
 \multicolumn{1}{c|}{0.8} &
 \multicolumn{1}{c|}{1} &
 1 &
 \multicolumn{1}{c|}{0.1} &
 \multicolumn{1}{c|}{0.08} &
 \multicolumn{1}{c|}{0.1} &
 0.06 &
 \multicolumn{1}{c|}{1} &
 1 \\ \hline
E5 &
 \multicolumn{1}{c|}{0.7} &
 \multicolumn{1}{c|}{0.47} &
 \multicolumn{1}{c|}{1} &
 0.69 &
 \multicolumn{1}{c|}{0.15} &
 \multicolumn{1}{c|}{0.08} &
 \multicolumn{1}{c|}{0.5} &
 0.29 &
 \multicolumn{1}{c|}{0.95} &
 1 \\ \hline
E6 &
 \multicolumn{1}{c|}{0.9} &
 \multicolumn{1}{c|}{1} &
 \multicolumn{1}{c|}{0.05} &
 1 &
 \multicolumn{1}{c|}{0.95} &
 \multicolumn{1}{c|}{1} &
 \multicolumn{1}{c|}{0.05} &
 0.08 &
 \multicolumn{1}{c|}{0.95} &
 1 \\ \hline
E7 &
 \multicolumn{1}{c|}{0.95} &
 \multicolumn{1}{c|}{0.41} &
 \multicolumn{1}{c|}{1} &
 0.74 &
 \multicolumn{1}{c|}{0.7} &
 \multicolumn{1}{c|}{0.44} &
 \multicolumn{1}{c|}{1} &
 0.51 &
 \multicolumn{1}{c|}{0.7} &
 1 \\ \hline
E8 &
 \multicolumn{1}{c|}{0.95} &
 \multicolumn{1}{c|}{0.42} &
 \multicolumn{1}{c|}{1} &
 0.69 &
 \multicolumn{1}{c|}{0.9} &
 \multicolumn{1}{c|}{0.51} &
 \multicolumn{1}{c|}{1} &
 0.5 &
 \multicolumn{1}{c|}{0.8} &
 1 \\ \hline
E9 &
 \multicolumn{1}{c|}{0.75} &
 \multicolumn{1}{c|}{0.94} &
 \multicolumn{1}{c|}{0.9} &
 0.95 &
 \multicolumn{1}{c|}{0.65} &
 \multicolumn{1}{c|}{0.41} &
 \multicolumn{1}{c|}{0.9} &
 0.47 &
 \multicolumn{1}{c|}{0.85} &
 0.77 \\ \hline
E10 &
 \multicolumn{1}{c|}{0.9} &
 \multicolumn{1}{c|}{0.72} &
 \multicolumn{1}{c|}{1} &
 0.59 &
 \multicolumn{1}{c|}{0.8} &
 \multicolumn{1}{c|}{0.46} &
 \multicolumn{1}{c|}{1} &
 0.51 &
 \multicolumn{1}{c|}{0.75} &
 0.88 \\ \hline
E14 &
 \multicolumn{1}{c|}{1} &
 \multicolumn{1}{c|}{1} &
 \multicolumn{1}{c|}{0.9} &
 0.9 &
 \multicolumn{1}{c|}{0.9} &
 \multicolumn{1}{c|}{0.45} &
 \multicolumn{1}{c|}{0.4} &
 0.44 &
 \multicolumn{1}{c|}{1} &
 1 \\ \hline
E15 &
 \multicolumn{1}{c|}{1} &
 \multicolumn{1}{c|}{1} &
 \multicolumn{1}{c|}{0.4} &
 0.84 &
 \multicolumn{1}{c|}{1} &
 \multicolumn{1}{c|}{0.49} &
 \multicolumn{1}{c|}{0.4} &
 0.42 &
 \multicolumn{1}{c|}{1} &
 1 \\ \hline
E16 &
 \multicolumn{1}{c|}{0.9} &
 \multicolumn{1}{c|}{0.86} &
 \multicolumn{1}{c|}{0.55} &
 1 &
 \multicolumn{1}{c|}{0.55} &
 \multicolumn{1}{c|}{0.58} &
 \multicolumn{1}{c|}{0.2} &
 0.12 &
 \multicolumn{1}{c|}{0.85} &
 1 \\ \hline
E17 &
 \multicolumn{1}{c|}{0.95} &
 \multicolumn{1}{c|}{0.97} &
 \multicolumn{1}{c|}{0.85} &
 0.94 &
 \multicolumn{1}{c|}{0.85} &
 \multicolumn{1}{c|}{0.63} &
 \multicolumn{1}{c|}{0.7} &
 0.93 &
 \multicolumn{1}{c|}{0.65} &
 1 \\ \hline
E20 &
 \multicolumn{1}{c|}{0.2} &
 \multicolumn{1}{c|}{1} &
 \multicolumn{1}{c|}{0.35} &
 1 &
 \multicolumn{1}{c|}{0.8} &
 \multicolumn{1}{c|}{0.97} &
 \multicolumn{1}{c|}{0.05} &
 0.2 &
 \multicolumn{1}{c|}{0.68} &
 1 \\ \hline
E21 &
 \multicolumn{1}{c|}{0.85} &
 \multicolumn{1}{c|}{0.97} &
 \multicolumn{1}{c|}{0.6} &
 1 &
 \multicolumn{1}{c|}{0} &
 \multicolumn{1}{c|}{0} &
 \multicolumn{1}{c|}{0.23} &
 0.43 &
 \multicolumn{1}{c|}{0.58} &
 1 \\ \hline
E23 &
 \multicolumn{1}{c|}{1} &
 \multicolumn{1}{c|}{0.96} &
 \multicolumn{1}{c|}{0.1} &
 1 &
 \multicolumn{1}{c|}{0.65} &
 \multicolumn{1}{c|}{0.62} &
 \multicolumn{1}{c|}{0} &
 0 &
 \multicolumn{1}{c|}{0.95} &
 1 \\ \hline
E24 &
 \multicolumn{1}{c|}{1} &
 \multicolumn{1}{c|}{0.88} &
 \multicolumn{1}{c|}{0} &
 0 &
 \multicolumn{1}{c|}{0.55} &
 \multicolumn{1}{c|}{0.68} &
 \multicolumn{1}{c|}{0} &
 0 &
 \multicolumn{1}{c|}{1} &
 1 \\ \hline
E25 &
 \multicolumn{1}{c|}{1} &
 \multicolumn{1}{c|}{0.94} &
 \multicolumn{1}{c|}{0} &
 0 &
 \multicolumn{1}{c|}{0.7} &
 \multicolumn{1}{c|}{0.64} &
 \multicolumn{1}{c|}{0} &
 0 &
 \multicolumn{1}{c|}{0.9} &
 0.95 \\ \hline
E26 &
 \multicolumn{1}{c|}{1} &
 \multicolumn{1}{c|}{0.96} &
 \multicolumn{1}{c|}{0} &
 0 &
 \multicolumn{1}{c|}{0.75} &
 \multicolumn{1}{c|}{0.6} &
 \multicolumn{1}{c|}{0} &
 0 &
 \multicolumn{1}{c|}{0.95} &
 0.91 \\ \hline
E27 &
 \multicolumn{1}{c|}{0.9} &
 \multicolumn{1}{c|}{0.9} &
 \multicolumn{1}{c|}{0.8} &
 0.62 &
 \multicolumn{1}{c|}{0.75} &
 \multicolumn{1}{c|}{0.72} &
 \multicolumn{1}{c|}{0.40} &
 0.39 &
 \multicolumn{1}{c|}{1} &
 1 \\ \hline
E28 &
 \multicolumn{1}{c|}{1} &
 \multicolumn{1}{c|}{0.95} &
 \multicolumn{1}{c|}{0.95} &
 0.47 &
 \multicolumn{1}{c|}{0.75} &
 \multicolumn{1}{c|}{0.76} &
 \multicolumn{1}{c|}{0} &
 0 &
 \multicolumn{1}{c|}{0.95} &
 1 \\ \hline
E29 &
 \multicolumn{1}{c|}{1} &
 \multicolumn{1}{c|}{0.9} &
 \multicolumn{1}{c|}{0.40} &
 0.48 &
 \multicolumn{1}{c|}{0.75} &
 \multicolumn{1}{c|}{0.64} &
 \multicolumn{1}{c|}{0.55} &
 0.54 &
 \multicolumn{1}{c|}{0.95} &
 1 \\ \hline
E30 &
 \multicolumn{1}{c|}{1} &
 \multicolumn{1}{c|}{0.43} &
 \multicolumn{1}{c|}{0.7} &
 0.56 &
 \multicolumn{1}{c|}{0.75} &
 \multicolumn{1}{c|}{0.69} &
 \multicolumn{1}{c|}{0.25} &
 0.19 &
 \multicolumn{1}{c|}{0.95} &
 0.85 \\ \hline
E31 &
 \multicolumn{1}{c|}{1} &
 \multicolumn{1}{c|}{0.42} &
 \multicolumn{1}{c|}{0.75} &
 0.56 &
 \multicolumn{1}{c|}{0.7} &
 \multicolumn{1}{c|}{0.6} &
 \multicolumn{1}{c|}{0.2} &
 0.17 &
 \multicolumn{1}{c|}{0.9} &
 0.91 \\ \hline
E32 &
 \multicolumn{1}{c|}{1} &
 \multicolumn{1}{c|}{0.83} &
 \multicolumn{1}{c|}{0.95} &
 0.83 &
 \multicolumn{1}{c|}{0.95} &
 \multicolumn{1}{c|}{0.47} &
 \multicolumn{1}{c|}{0.95} &
 0.46 &
 \multicolumn{1}{c|}{1} &
 1 \\ \hline
E33 &
 \multicolumn{1}{c|}{1} &
 \multicolumn{1}{c|}{0.56} &
 \multicolumn{1}{c|}{1} &
 0.91 &
 \multicolumn{1}{c|}{1} &
 \multicolumn{1}{c|}{0.49} &
 \multicolumn{1}{c|}{1} &
 0.48 &
 \multicolumn{1}{c|}{1} &
 1 \\ \hline
Average &
 \multicolumn{1}{c|}{0.91} &
 \multicolumn{1}{c|}{0.82} &
 \multicolumn{1}{c|}{0.65} &
 0.72 &
 \multicolumn{1}{c|}{0.65} &
 \multicolumn{1}{c|}{0.52} &
 \multicolumn{1}{c|}{0.39} &
 0.3 &
 \multicolumn{1}{c|}{\textbf{0.89}} &
 \textbf{0.96} \\ \hline
\end{tabular}

\caption{Event detection performance under the single-target and multi-target scenarios. The results of WiFinger under two scenarios are the same and thus merged.}
\label{tab:regular}
\end{table*}

\begin{figure}
 \centering
 \includegraphics[width=.7\linewidth]{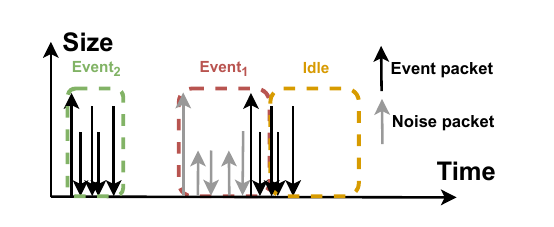}
 \caption{The influence of mismatch in continuous event tracking. The model skips half of the $Event_2$ packets for misclassifying $Event_1$.}
 \label{fig:mis_penalty}
\end{figure}

In the most advanced and informative scenario, the performance gap becomes event more significant. WiFinger maintains the highest recall and precision rates of 86\% and 95\%, while Peek-a-boo and IoTBeholder now only achieve 49\% and 46\% recall rates, and 48\% and 35\% precision rates respectively, barely useable.\footnote{Considering that 30 samples may not be sufficient for training robust ML-based models, we further collected 100 training samples for each event, but observe no performance improvement.} Such results reflect the models' authentic performance on real-world tracking that have been overrated in previous evaluations. Most of the existing works evaluate classification performances on chunked samples, where the results of different samples are classified independently. Nonetheless, in the multi-target scenario, misclassifications have impact on subsequent detections such as overlooking other events, as shown in Figure~\ref{fig:mis_penalty}. Therefore, during event tracking, false positive detections have a much higher impact on the actual overall performance. Among the three models, IoTBeholder achieves the worst performance. IoTBeholder uses the confidence score of a group of binary classifier for multi-class classification. Therefore, once the testing data deviates from the training data pattern, the corresponding model of events can no longer output a dominating confidence score and easily lead to erroneous judgment. As a comparison, WiFinger outstands for its excellent precision rate. Due to the event-unique base fingerprint features, one event hardly gets matched to a different fingerprint. Furthermore, we experimented with matching fingerprints across devices, and their uniqueness demonstrates the capability of facilitating simultaneous device fingerprinting and event identification. Such cross-device uniqueness has also been verified in previous works~\cite{pingpong2019trimananda,wan2021iotathena}.

\subsection{Robustness Analysis}
\subsubsection{Sensitivity Against Noises}
\label{sec:sensitivity}
Despite promising overall performances, WiFinger performs considerably low on the recall rates of certain events. We manually analyze the matching results of FNs and FPs to analyze the main causes. \textbf{FN:} WiFinger drops a detection result either for not matching enough packets of the base fingerprint or the failure of aligning subsequences temporally, where the former situation happens much more frequently. In our parameter settings, a successful match must comprise 60\% ($\gamma$) of the base fingerprint (except Mi Sweeper).  The choice of  $\gamma$ is due to the bursty drop pattern of wireless sniffing, that an average loss rate of around 10-20\% may cause some events having 30-40\% missing packets. Moreover, for short base fingerprints of only 2-4 packets, even the ``60\%'' matching requirement poses a significant challenge. Given the brevity of short fingerprints, each packet carries a disproportionately large weight, meaning a single mismatch can severely undermine the matching percentage. For example, the fingerprint of E17 (Mi Sweeper) only involves 2 packets, indicating that missing any packet leads to the failure of meeting the ``60\%'' requirement. \textbf{FP:} we also notice that packet loss is the main reason for misclassifying events with similar base fingerprints. For example, E7 and E8 adjust the volume of the Google Home speaker, and a few fingerprint packets of E7 are a byte larger than E8. Nevertheless, if such distinguishing packets are lost from E8, its remaining packets could be perfectly mismatched as an E7 event. Such mismatches also happen to some idle states whose traffic exhibits similar patterns. For instance, idle traffic of Mi Sweeper can occasionally match 50\% or 75\% of the E16 fingerprint. Diving deeper into the situation, we notice that such special phenomena do not happen to any other devices/events and only occurs after the ``stop sweeping'' commands. A reasonable explanation is that there exists hidden events happening to share similar traffic patterns, e.g., reporting the ``back-to-charging'' status.

In summary, short base fingerprints are much more sensitive to packet losses, causing the majority of the FNs and FPs. Nonetheless, depending on the needs, attackers can adjust the matching percentage parameter for these events to balance the trade-off between precision and recall. If attackers increase the matching percentage $\gamma$, they could obtain higher precision but lower recall, or vice versa. In later sections, we investigate the choice of parameters more thoroughly.

\subsubsection{Sensitivity against Channel Variations}
\begin{figure*}[htb]
 \centering
 \begin{subfigure}{.19\linewidth}
  \centering
  \includegraphics[width=\linewidth]{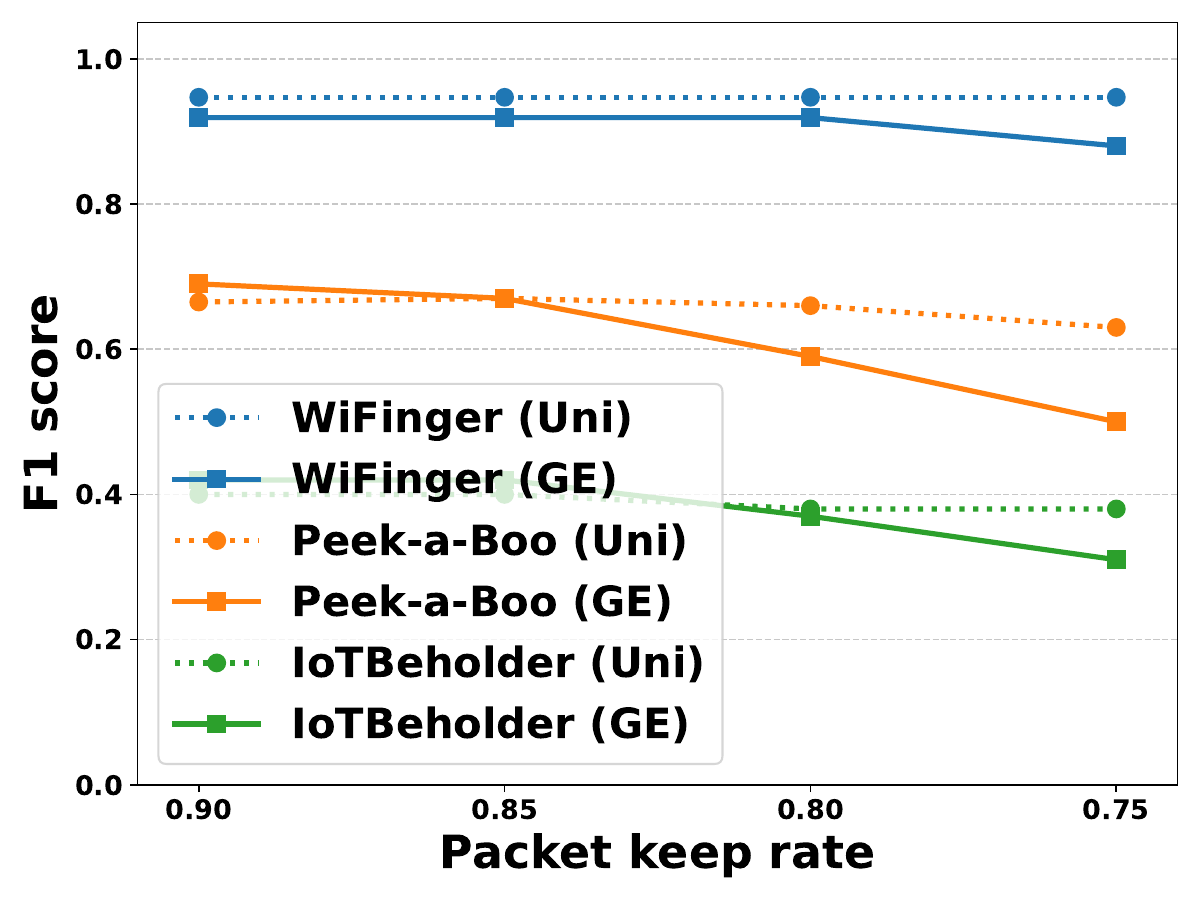}
 \caption{Gosund Plug (B)}
 \end{subfigure}
 \begin{subfigure}{.19\linewidth}
 \centering
 \includegraphics[width=\linewidth]{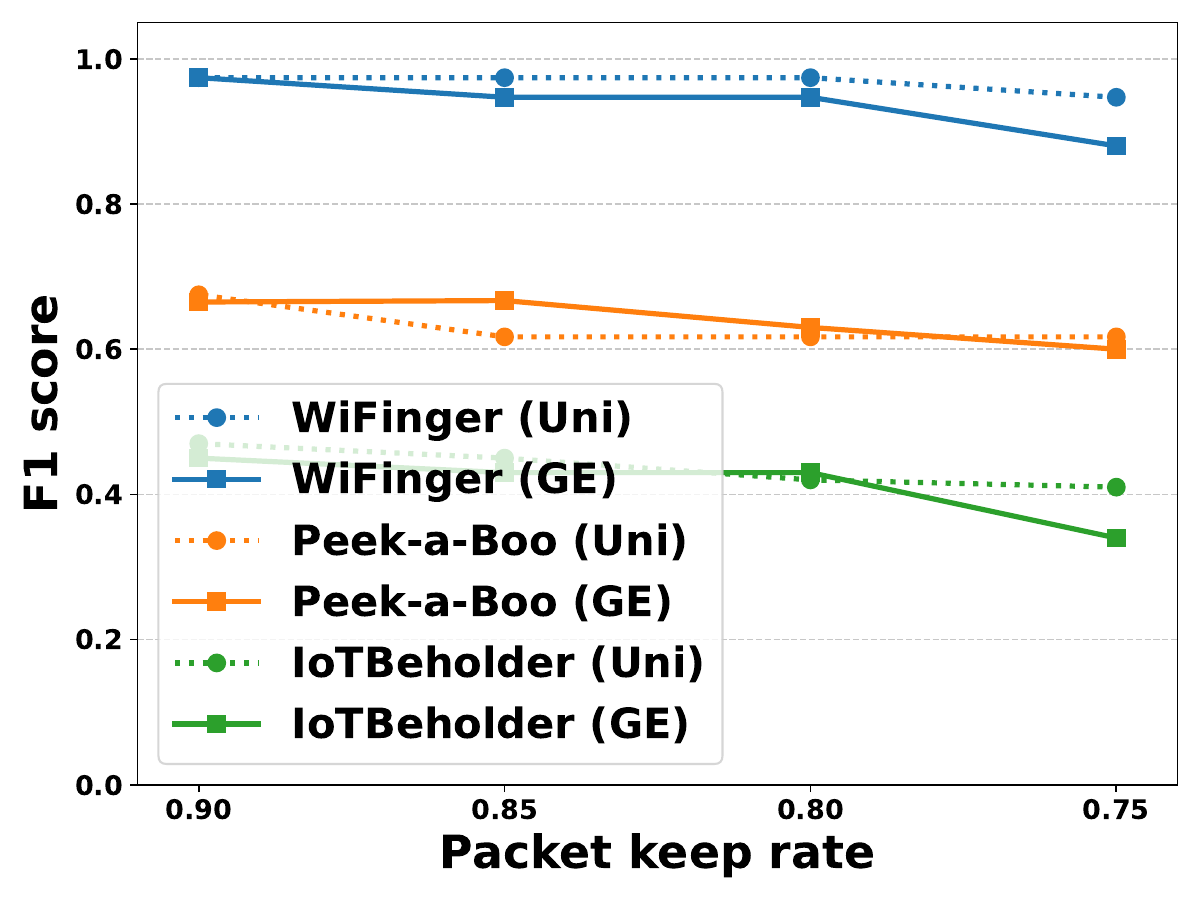}
 \caption{ICX AC (B)}
 \end{subfigure}
 \begin{subfigure}{.19\linewidth}
 \centering
 \includegraphics[width=\linewidth]{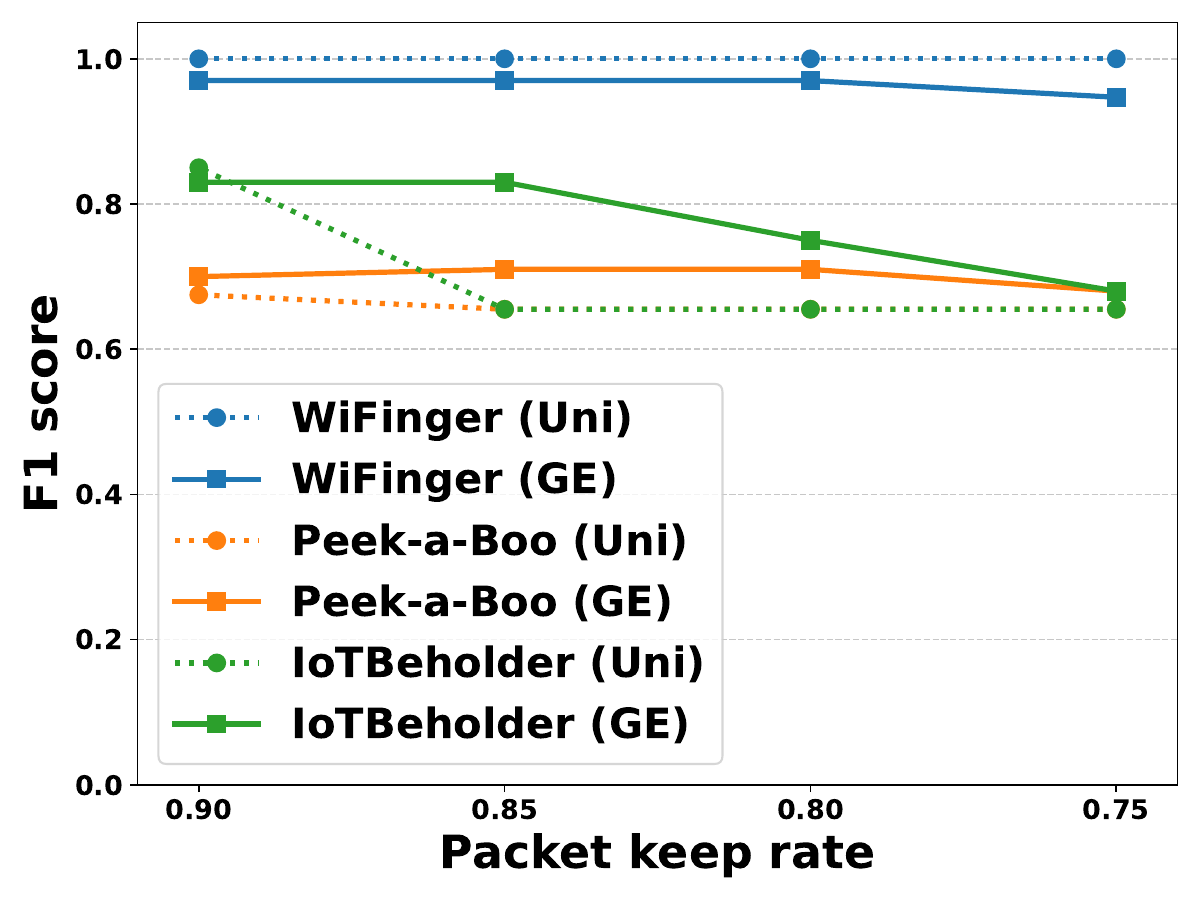}
 \caption{TP-Link Plug (B)}
 \end{subfigure}
 \begin{subfigure}{.19\linewidth}
 \centering
 \includegraphics[width=\linewidth]{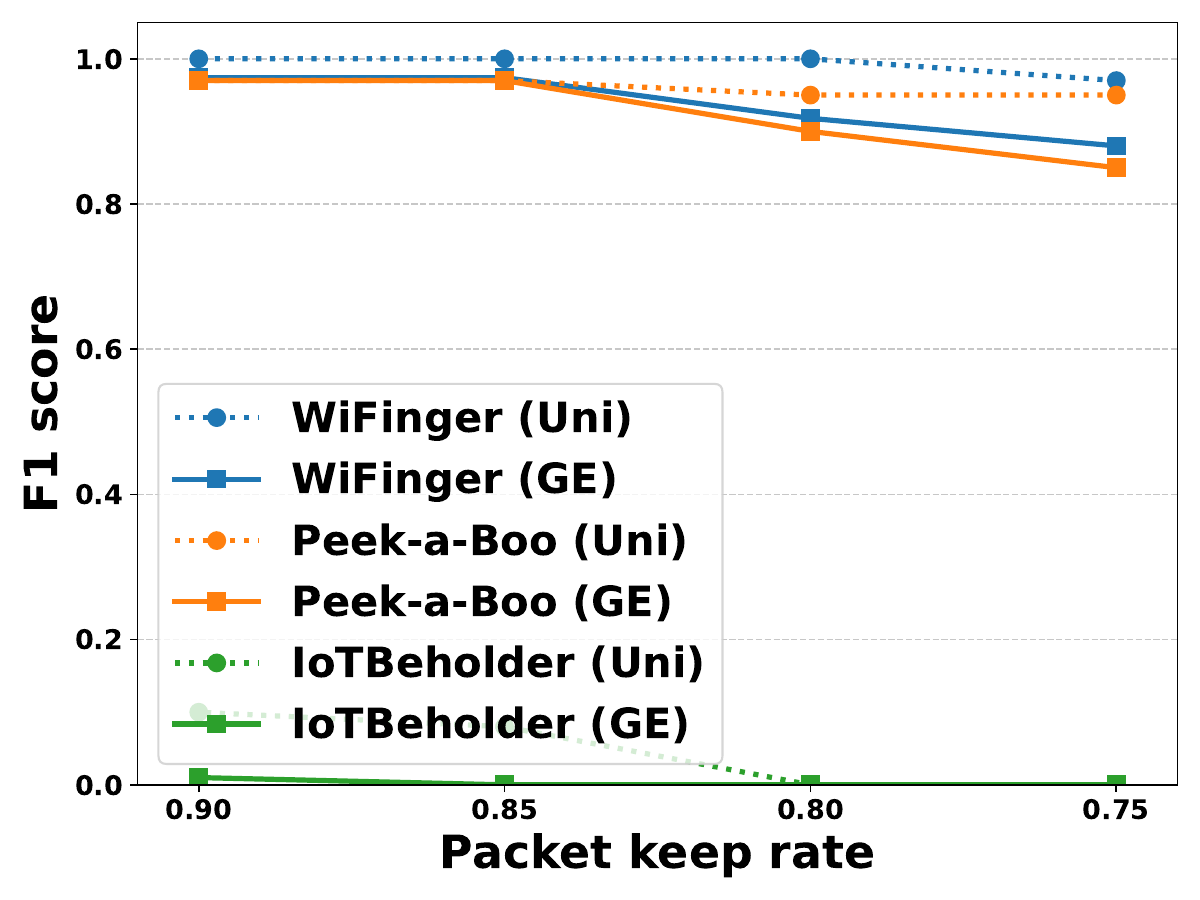}
 \caption{Alexa Time (ST)}
 \end{subfigure}
 \begin{subfigure}{.19\linewidth}
 \centering
 \includegraphics[width=\linewidth]{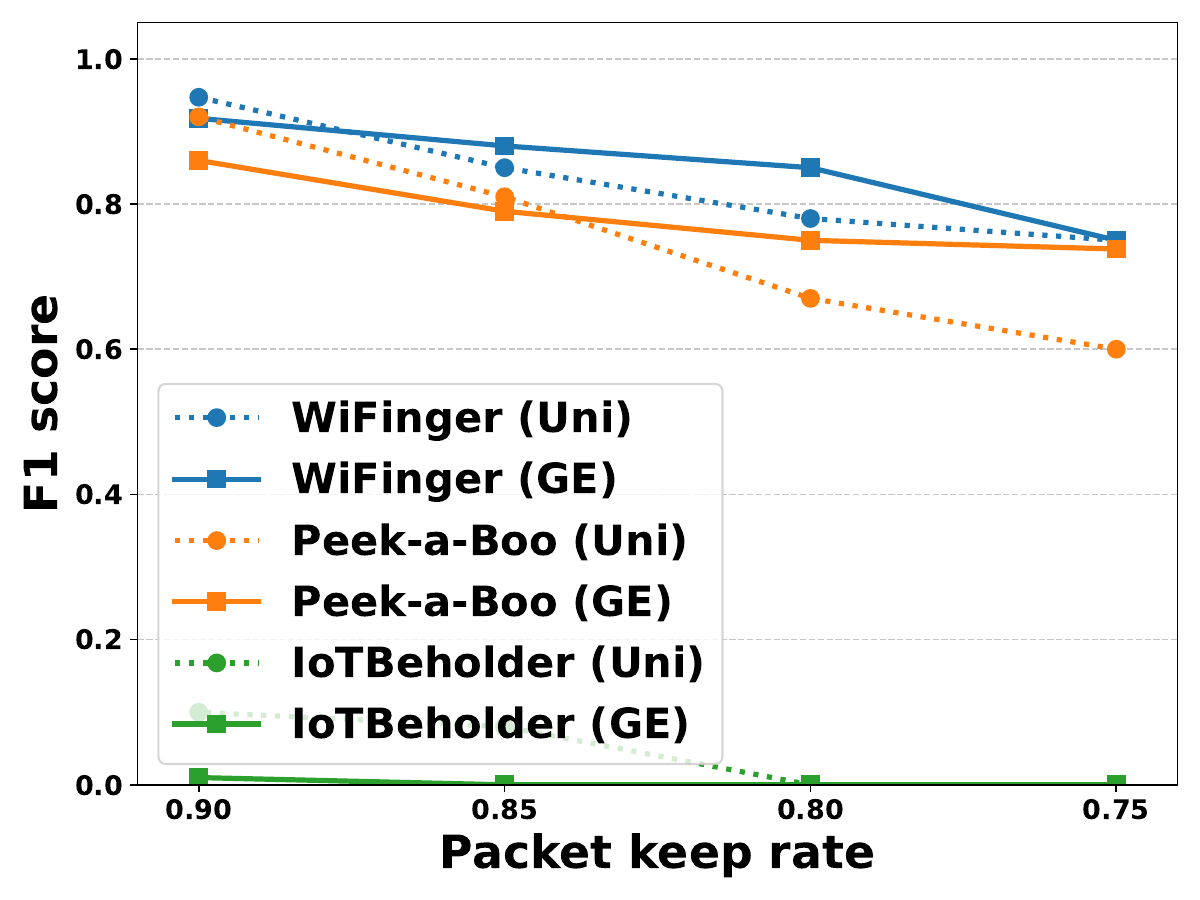}
 \caption{Alexa DND (ST)}
 \end{subfigure}
 \caption{Average model performances under different packet keep rates. B stands for ``binary'', ST stands for ``single-target.''}
 \label{fig:various_packet_loss}
\end{figure*}

During the attack, wireless channel conditions largely impact the fingerprinting quality. To evaluate its influences on fingerprints with various lengths, we select 5 events with the highest keep rate (\textgreater 90\%), discard some data packets to reach predefined packet loss rates, and evaluate models under worse ``simulated'' conditions. We use a uniform packet drop as the baseline and the well-established Gilbert-Elliott (GE)~\cite{konrad2001markov,hasslinger2008gilbert} model to simulate more realistic bursty packet loss of wireless channels. We repeat each process three times to mitigate abnormality. As shown in Figure.~\ref{fig:various_packet_loss}, WiFinger consistently outperforms Peek-a-Boo and IoTBeholder under both simulation model, demonstrating robustness even under various packet loss rates. 
Additionally, we note that both IoTBeholder and Peek-a-Boo demonstrate some level of robustness against packet losses, despite their worse overall performance. This indicates that burst-level ML analysis are still capable of distinguishing very unique events from idle states, but their effectiveness largely depend on the nature of events.

\subsection{Limits and Effectiveness: Ablation Study on Parameters}
\begin{figure}
 \centering
 \begin{subfigure}{.49\linewidth}
  \centering
  \includegraphics[width=\linewidth]{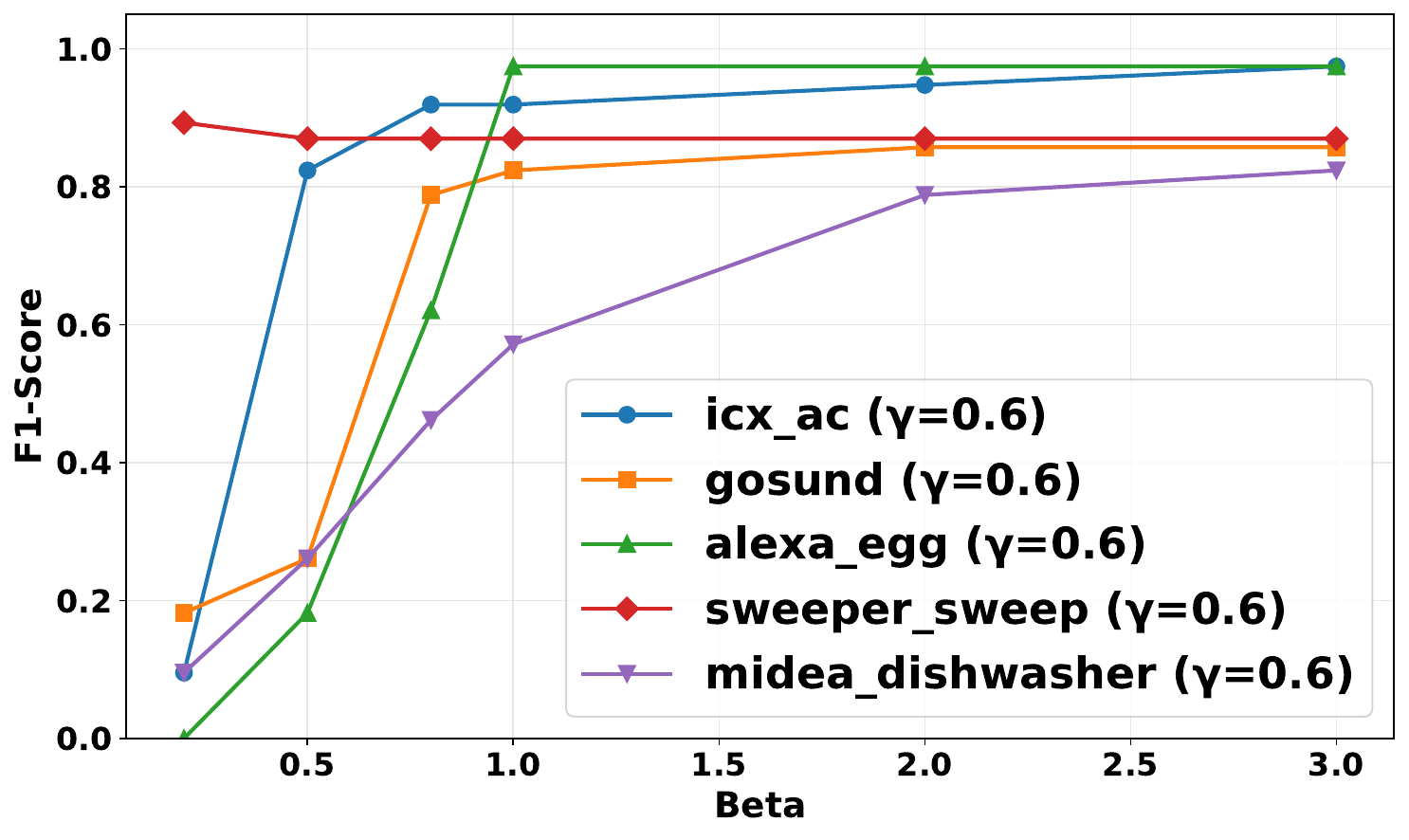}
  \caption{F1-score v.s. $\beta$}
 \end{subfigure}
 \begin{subfigure}{.49\linewidth}
  \centering
  \includegraphics[width=\linewidth]{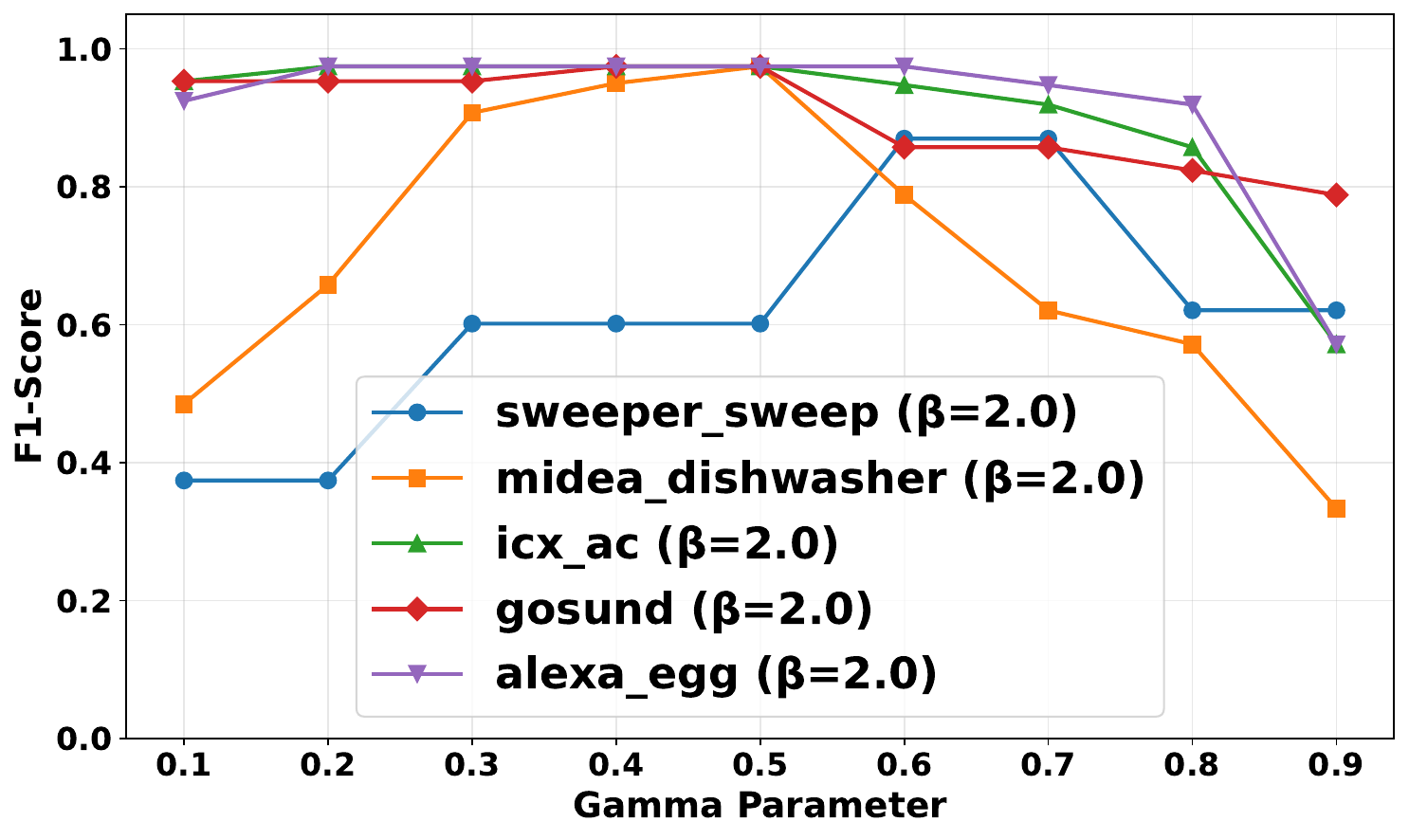}
  \caption{F1-score v.s. $\gamma$}
 \end{subfigure}
 \caption{Ablation study on parameter $\beta$ and $\gamma$.}
 \label{fig:ablation}
\end{figure}

According to existing analysis, WiFinger's performance is determined by three factors: sniffing data quality, fingerprint length, and parameter settings. While the former two factors are unadjustable after selecting the target and setting up the sniffer, attackers could adapt parameters to their need on precision and recall rates, depending on the current sniffing channel condition and their target fingerprints. Therefore, understanding a balanced setting and the boundary of effectiveness is crucial for deploying the attack. WiFinger has three adjustable parameters: similar packet size gap $\epsilon$, minimum sequence matching percentage $\gamma$, and interval distance threshold $\beta$. During the experiment, we empirically set $\epsilon$ as 1, meaning two packets are only considered similar if they have the exact same sizes and transmission directions. The influence of $\beta$ and $\gamma$ is shown in Figure~\ref{fig:ablation}.

Increasing $\beta$ allows WiFinger to accommodate more network jitters during the transmission. WiFinger reaches a stable performance when $\beta$ is larger than 2 seconds. Generalizing to various network conditions, setting $\beta=3$ is sufficiently accommodating. Additionally, WiFinger achieves best performance when $\gamma$ is around 0.5-0.6. There are several factors influencing the value. First of all, $\gamma$ depends on the sniffing data quality. In an extreme case where all packets could be sniffed and the network is stable, attackers would expect matching the whole base fingerprint every single time, i.e., $\gamma=1$. Yet, when a distinctive packet differentiating an event from idle traffic is lost (e.g., \textit{event}: \{155U, 255D, 137U\}; \textit{idle}: \{155U, 255D\}), distinguishing events is the same as making a random guess. To encounter such inevitable scenarios, attackers could adjust $\gamma$ to balance the trade-off on precision and recall. Higher $\gamma$ suggests higher matching percentage of the fingerprint. If attackers seek for capturing events with high confidence for subsequent actions (as implied in Section~\ref{sec:security_impact}), tuning $\gamma$ towards a higher value provides more guarantees, and vice versa. Nonetheless, either too high or low $\gamma$ results in the increment of false negatives or false positives, lowering the overall performances (F1-score). As a result, we suggest set $\gamma$ as 0.1\TILDE0.2 below $loss_{sniff}$ a balanced choice, and adjust it according to actual sniffing capability and the length of fingerprints. Lower $\gamma$ may work, but only for very unique and long fingerprints.

\subsection{Events and Fingerprint Case Study}
\label{sec:case_study}
Table~\ref{tab:events} lists the events and devices used in the experiments. We demonstrate the metadata of the extracted base fingerprints in Table~\ref{tab:metadata}. First, complex devices may have very different fingerprint lengths and durations, depending on the commands' complexity. Alexa Echo's fingerprints' (E1-E6) lengths vary from 4 to 41 packets, and the duration varies from 0.2s to 7.7s. Second, correlated events mostly have very similar pattern. For instance, events in E11 (or E12-E15) have exactly the same fingerprints, i.e., same fingerprint lengths, packet sizes, directions, and similar interval distributions. These events are inseparable from encrypted traffic analysis. Nevertheless, some correlated events share subtle differences: on/off commands of the Wiz Hue Light (E9-E10) have single-byte differences on their first two fingerprint packets; the volume up/down commands of the Google Home (E7-E8) also have single-byte differences on several fingerprint packets, but the two commands also have different lengths as well. Last but not least, complex events traffic typically involve more sub-bursts (segments) during their span. These sub-bursts increase the difficulty of burst-classification or window-selection for ML-based approaches, but turn out to be useful for handling large volume of traffic with our segmentation technique. In general, IoT fingerprints have both significant flow-level differences and subtle packet-level variations, making it harder for tuning ML models to capture their traffic patterns. 

\begin{table}[h!]
\centering
\setlength{\tabcolsep}{3pt} 
\renewcommand{\arraystretch}{1} 
\begin{tabular}{|c||c|c|c|}
\hline
Event ID & Packet Num & Duration(s) & Sub-bursts \\ \hline
E1  & 4      & 0.23        & 1          \\ \hline
E2  & 23     & 4.5         & 3          \\ \hline
E3  & 41     & 7.7         & 6          \\ \hline
E4  & 28     & 3.15        & 3          \\ \hline
E5  & 42     & 4.03        & 4          \\ \hline
E6  & 16     & 0.35        & 1          \\ \hline
E7  & 9      & 3.29        & 2          \\ \hline
E8  & 17     & 3.41        & 3          \\ \hline
E9  & 4      & 0.5         & 1          \\ \hline
E10 & 4      & 0.5         & 1          \\ \hline
E11 & 4      & 0.2         & 1          \\ \hline
E12 & 2      & 0.04        & 1          \\ \hline
E13 & 20     & 1.52        & 2          \\ \hline
E14 & 11     & 0.62        & 1          \\ \hline
E15 & 10     & 0.81        & 1          \\ \hline
E16 & 4      & 5.69        & 2          \\ \hline
E17 & 2     & 5.76        & 2          \\ \hline
E18 & 15     & 1.25        & 1          \\ \hline
E19 & 6     & 0.42        & 1          \\ \hline
E20 & 4     & 0.73        & 1          \\ \hline
E21 & 4     & 0.51        & 1          \\ \hline
E22 & 7     & 1.13        & 2          \\ \hline
E23 & 10     & 3.81        & 2          \\ \hline
E24 & 9     & 4.24        & 2          \\ \hline
E25 & 13     & 3.75        & 2          \\ \hline
E26 & 11     & 3.44        & 2          \\ \hline
E27 & 8     & 6.91        & 2          \\ \hline
E28 & 8     & 6.85        & 2          \\ \hline
E29 & 8     & 6.93        & 2          \\ \hline
E30 & 11     & 6.87        & 3          \\ \hline
E31 & 13     & 6.83        & 3          \\ \hline
E32 & 11     & 0.88        & 1          \\ \hline
E33 & 10     & 0.85        & 1          \\ \hline
\end{tabular}
\caption{Metadata of the extracted base fingerprints.}
\label{tab:metadata}
\end{table}

\subsection{Efficiency Analysis}
We introduced anchor reference and fingerprint segmentation to optimize the efficiency and accuracy of FMLCS for large-volume traffic devices (E1-E8). We use the multi-target setting to compare their online matching performance. Each testing file contains 20 events spanning over 1200 seconds. As shown in Table~\ref{tab:afmlcs_comparison}, AFMLCS requires significantly less time to process the traffic, and the time cost difference increases with traffic volume. For example, during a sudden burst of traffic exceeding one hundred seconds in Event 5 (E5), FMLCS spends most of the time handling the burst but exhibits even worse performance.
Comparing with existing approaches, we measure the average time for process all testing sequences with the three approaches. As shown in Table~\ref{tab:efficiency_comparison}, WiFinger outperforms with approximate 10x processing speed.

Moreover, AFMLCS also plays a very important role in the base fingerprint extraction. During the extraction, WiFinger applies FMLCS to extract consensus sequences pairwisely among all event traffic bursts. Yet, the matching processes between two noisy sequences make baseline FMLCS stuck at the DP function for its exponential time complexity. As a result, FMLCS could not finish the process for E2-E8 even after hours of waiting, while AFMLCS only takes less than a minute to do so. In general, for short fingerprints/traffic where segmentation techniques are not applicable, WiFinger degrades AFMLCS to the baseline version for its simplicity and effectiveness.

\begin{table}[h]
\centering
\begin{tabular}{|c|ccc|}
\hline
\multirow{2}{*}{Traffic Volume} & \multicolumn{3}{c|}{Time Cost (s)}                                            \\ \cline{2-4} 
                                & \multicolumn{1}{c|}{Peek-a-Boo} & \multicolumn{1}{c|}{IoTBeholder} & WiFinger \\ \hline
Small                           & \multicolumn{1}{c|}{0.53}       & \multicolumn{1}{c|}{0.35}        & 0.04     \\ \hline
Large                           & \multicolumn{1}{c|}{5.18}       & \multicolumn{1}{c|}{8.1}         & 0.45     \\ \hline
\end{tabular}
\caption{Average time cost for processing testing packet sequence of all events.}
\label{tab:efficiency_comparison}
\end{table}

\begin{table}[]
\centering
\setlength{\tabcolsep}{3pt} 
\renewcommand{\arraystretch}{1.05} 
\begin{tabular}{|c|c|cc|cc|}
\hline
\multirow{2}{*}{Event ID} & \multirow{2}{*}{Packet Num} & \multicolumn{2}{c|}{Process Time(s)} & \multicolumn{2}{c|}{F1 Score} \\ \cline{3-6} 
      &       & \multicolumn{1}{c|}{FMLCS} & AFMLCS & \multicolumn{1}{c|}{FMLCS} & AFMLCS \\ \hline
E1    & 3854  & \multicolumn{1}{c|}{0.13}  & 0.13   & \multicolumn{1}{c|}{0.85}  & 0.80   \\ \hline
E2    & 5740  & \multicolumn{1}{c|}{0.17}  & 0.07   & \multicolumn{1}{c|}{1.00}  & 1.00   \\ \hline
E3    & 8178  & \multicolumn{1}{c|}{2.34}  & 0.15   & \multicolumn{1}{c|}{0.92}  & 0.97   \\ \hline
E4    & 12670 & \multicolumn{1}{c|}{7.97}  & 0.85   & \multicolumn{1}{c|}{0.97}  & 1.00   \\ \hline
E5    & 18011 & \multicolumn{1}{c|}{39.2}  & 1.3    & \multicolumn{1}{c|}{0.33}  & 0.97   \\ \hline
E6    & 1789  & \multicolumn{1}{c|}{0.14}  & 0.15   & \multicolumn{1}{c|}{1.00}  & 0.97   \\ \hline
E7/E8 & 4179  & \multicolumn{1}{c|}{0.34}  & 0.1    & \multicolumn{1}{c|}{0.84}  & 0.86   \\ \hline
\end{tabular}
\caption{AFMLCS improves both efficiency and accuracy for large-volume traffic events.}
\label{tab:afmlcs_comparison}
\end{table}

\subsection{Countermeasures}
\label{sec:countermeasures}
\begin{figure}
 \centering
 \includegraphics[width=1\linewidth]{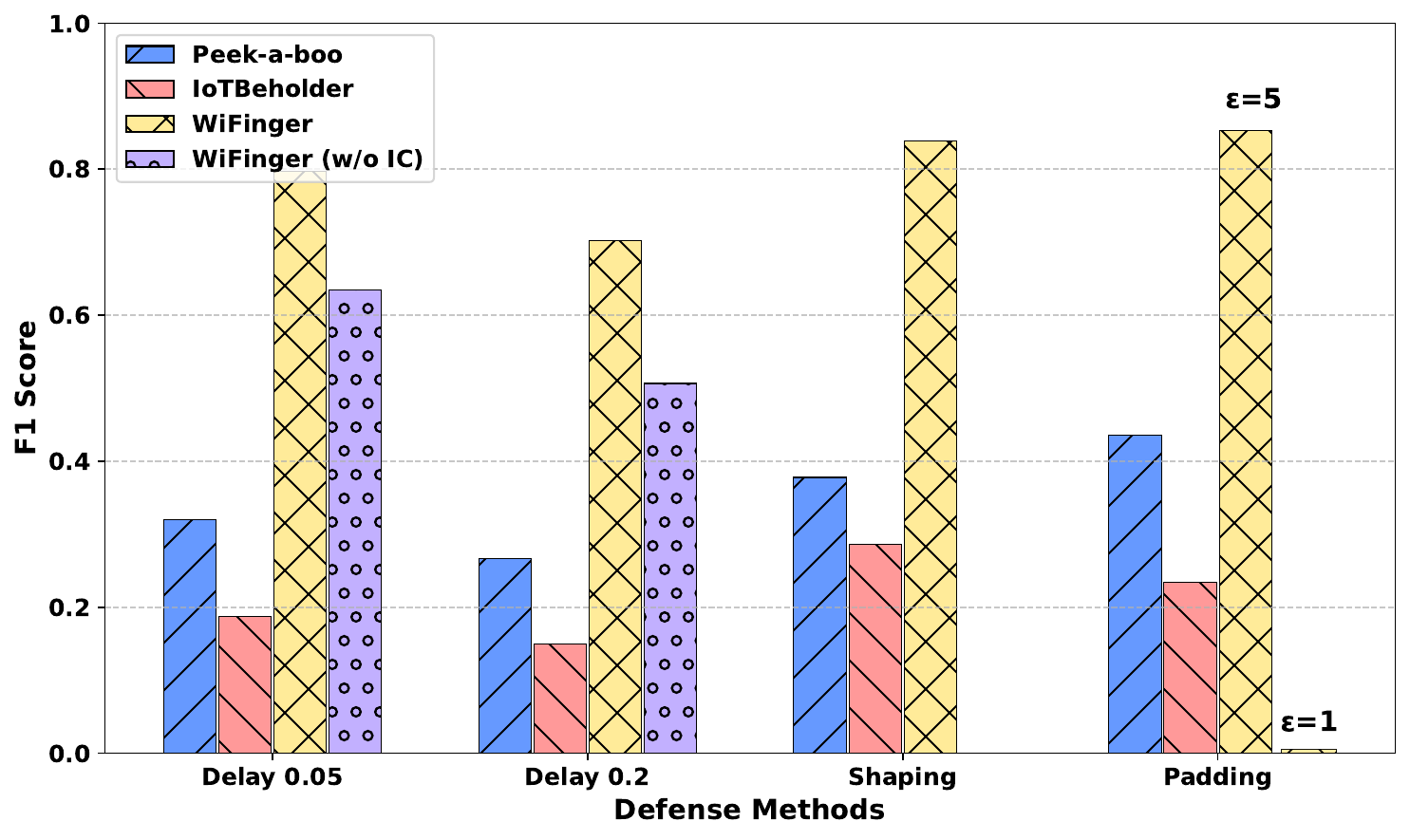}
 \caption{The three methods' average attacking performances against different defenses. Ablation study on IC is not conducted for the latter two due to their irrelevance.}
 \label{fig:defense}
\end{figure}

To protect users from such privacy inference attacks, previous works have proposed several countermeasures, including traffic shaping~\cite{apthorpe2018keepingprivate}, traffic delaying~\cite{cai2014cs}, and packet padding~\cite{dyer2012peek,cai2014systematic}.  Traffic delaying delays packet transmission for a random short period of time to obfuscate the packet intervals. Knowledgeable defenders can also use it to crack WiFinger by enlarge packet intervals beyond the threshold requirements. Traffic shaping randomly inserts dummy packets in both direction to masquarade the flow-level characteristics. Packet padding adds dummy bytes to each packet to obfuscate the size of payload. To emulate these protection mechanisms, we add random delays (0-0.05s or 0-0.2s) to packets for traffic delaying, add dummy packets to the original traffic for traffic shaping, and slightly increase all packet sizes (1-5 bytes randomly) for packing padding defense, respectively. We evaluate current fingerprints against the three defenses under the multi-target setting, and the effectiveness of these countermeasures against WiFinger is shown in Figure~\ref{fig:defense}. 

First, traffic shaping has a very limited influence on the WiFinger. This is because dummy packets influence neither the intervals between fingerprint packets nor their sizes and directions. Consequently, the shaped traffic still embodies the same event patterns, and thus most events end up with a similar performance with a slight drop of the recall rate. Its only impact is that it increases the processing time of complicated events at the cost of a large bandwidth. 

Traffic delay also has a limited impact on WiFinger. During the defense, transmission delays will be accumulated for each packet, i.e., if a request is delayed, its response will be also delayed for the same amount of time plus its own delay. Consequently, the intervals between tail packets and header packets are expected to increase, even exceeding the temporal alignment threshold of FMLCS. Nonetheless, thanks to our proposed IC mechanism, WiFinger clips and calibrates out-of-distribution intervals to limit the delay accumulation effect, ensuring temporal alignments between fingerprints and new sequences. As shown in the ablation study in Figure~\ref{fig:defense}, WiFinger without IC performs dramatically worse under similar delaying defenses.

Traffic padding demonstrates to be the most effective defense against WiFinger. Due to the direct change of packet sizes, the examination process (FMLCS) on packet sizes and directions is destroyed. As a result, the recall and precision rates drop to 0\% for all events if WiFinger still keeps $\epsilon=1$. Nonetheless, as we adapt WiFinger to the defense by loosing $\epsilon$ to 5 bytes, the performance of WiFinger recovers to a degree as if no defense is applied. For a padding defense to be effective, it should add enough dummy bytes (e.g., $\geq 100$) to obfuscate packets of different types.

\smallskip
\noindent\textbf{Cost-Effectiveness tradeoff:} Since WiFinger aims at the first-hand wireless sniffing, protections must be implemented on the IoT devices, which inevitably increases the burden of IoT application developers and manufactures. Moreover, the three defenses come at different costs on transmission speed or bandwidth. 

\begin{figure}[h!]
  \centering
  \includegraphics[width=\linewidth]{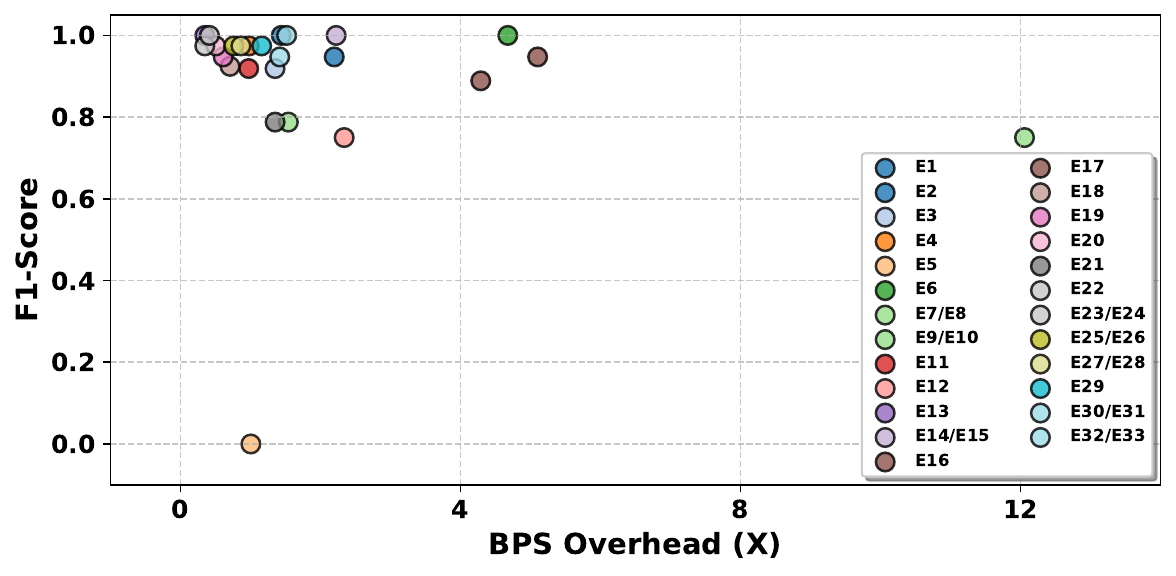}
  \caption{Additional bandwidth overhead (\# times of bps) introduced by traffic shaping defense.}
  \label{fig:shape_cost}
\end{figure}

\textbf{Traffic shaping} (Fig.~\ref{fig:shape_cost}) takes more bandwidth for the significant amount of extra dummy packets to be transmitted. Transmission rates of events are several times (from two to over ten times) higher than their regular rate, causing unacceptable overhead.

\begin{figure}[h!]
  \centering
  \includegraphics[width=\linewidth]{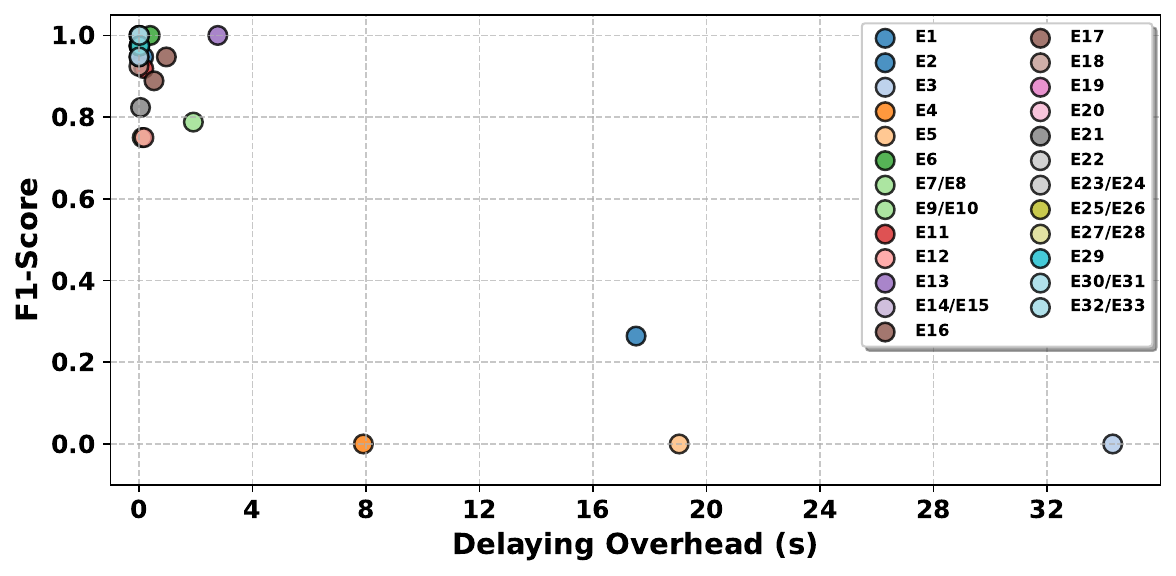}
  \caption{Additional event execution overhead (seconds) introduced by traffic delaying defense.}
  \label{fig:delay_cost}
\end{figure}

\textbf{Traffic delaying} (Fig.~\ref{fig:delay_cost}) increases the latency of an event execution and demonstrates extremely biased performance among events. On the one hand, they perform well for large-volume traffic events, but all events' executions are delayed by 7 to 30 seconds, severely impacting the system's utility. On the other hand, for small-volume traffic, while the accumulated delays of 0.5-1 seconds do not impact utility, they are also not sufficient for protecting traffic from WiFinger attacks.

\begin{figure}[h!]
  \centering
  \includegraphics[width=\linewidth]{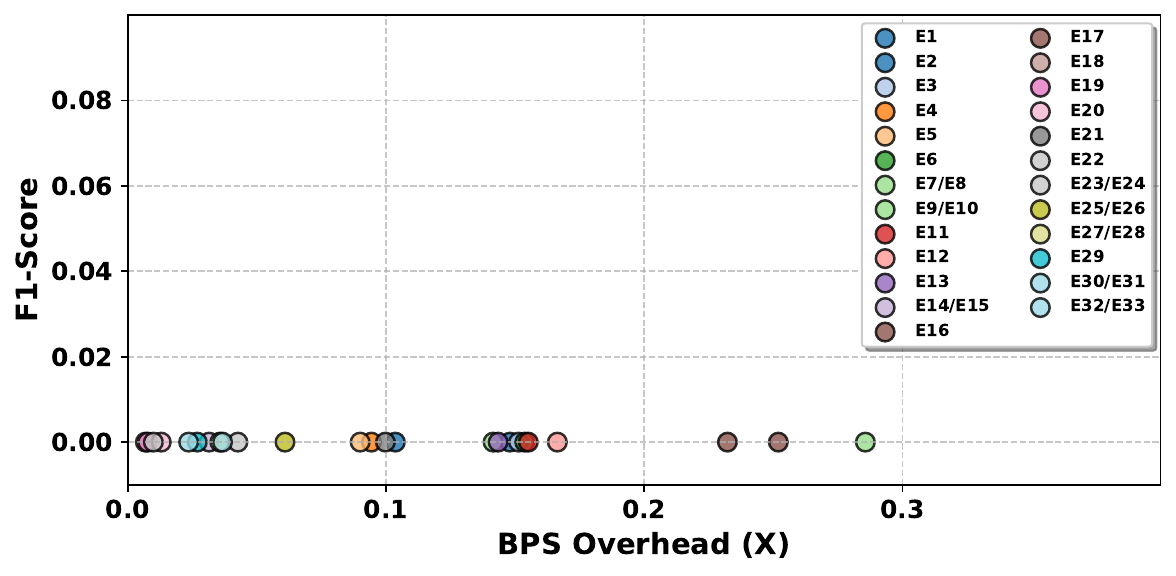}
  \caption{Additional bandwidth overhead (\# times of bps) introduced by packet padding defense.}
  \label{fig:padding_cost}
\end{figure}

\textbf{Packet padding} (Fig.~\ref{fig:padding_cost}) applies relative small packet-level overheads compared to the other two, but achieves strong effects on breaking fingerprinting approaches relying on the size of the payload. We tested randomly padding 1-100 dummy bytes to packets and successfully disabled WiFinger at the cost of around 20\% overhead on the transmission bandwidth. Therefore, packet padding works well as the first line of defenses against WiFinger and all other packet-matching attacks. Some valid implementation options of randomizing the payload sizes include using OkHttp~\cite{okhttp} to pad HTTP headers or using padding functionalities of the middleware protocols such as TLS. Since existing defenses may only support web development, developers still need to integrate them to the embedded IoT devices.

\label{app:event_id}
\begin{table*}[h]
    \centering
\begin{tabular}{|c|c|c|c|}
\hline
ID     & Device Type                                & Device                             & Related Events                    \\ \hline
E1     & \multirow{8}{*}{Agentic Controller}        & \multirow{6}{*}{Alexa Echo Dot}    & DND/UnDND                         \\ \cline{1-1} \cline{4-4} 
E2 (H) &                                            &                                    & Q1: What time is it?              \\ \cline{1-1} \cline{4-4} 
E3 (H) &                                            &                                    & Q2: What's the price of eggs?     \\ \cline{1-1} \cline{4-4} 
E4 (H) &                                            &                                    & Q3: What's the weather now?       \\ \cline{1-1} \cline{4-4} 
E5 (H) &                                            &                                    & Q4: What's the weather like in X? \\ \cline{1-1} \cline{4-4} 
E6     &                                            &                                    & Volume Up/Down                    \\ \cline{1-1} \cline{3-4} 
E7 (H) &                                            & \multirow{2}{*}{Google Home}       & Volume Up                         \\ \cline{1-1} \cline{4-4} 
E8 (H) &                                            &                                    & Volume Down                       \\ \hline
E9     & \multirow{7}{*}{Smart Peripherals}         & \multirow{2}{*}{Wiz Hue Light}     & On                                \\ \cline{1-1} \cline{4-4} 
E10    &                                            &                                    & Off                               \\ \cline{1-1} \cline{3-4} 
E11    &                                            & TP-Link Plug                       & On/Off                            \\ \cline{1-1} \cline{3-4} 
E12    &                                            & ICX-RF A/C Controller              & On/Off                            \\ \cline{1-1} \cline{3-4} 
E13    &                                            & Gosund Plug                        & On/Off                            \\ \cline{1-1} \cline{3-4}
E14    &                                            & \multirow{2}{*}{WAH Plug}                        & On                            \\ \cline{1-1} \cline{4-4}
E15    &                                            &                                    & Off                               \\ \hline
E16    & \multirow{19}{*}{Integrated Smart Actuator} & \multirow{2}{*}{Mi Sweeper}        & On/Off                            \\ \cline{1-1} \cline{4-4} 
E17    &                                            &                                    & Mode Silent/Standard/Strong       \\ \cline{1-1} \cline{3-4} 
E18    &                                            & Midea Dishwasher                   & On/Off                            \\ \cline{1-1} \cline{3-4} 
E19    &                                            & Midea Dish Sterilizer               & On/Off                            \\ \cline{1-1} \cline{3-4} 
E20    &                                            & \multirow{2}{*}{Xiaomi Humidifier} & On/Off                            \\ \cline{1-1} \cline{4-4} 
E21    &                                            &                                    & Continuous humidification/Close   \\ \cline{1-1}\cline{3-4}
E22    &                                            & Wi-Fi Sprinkler               & On/Off                            \\ \cline{1-1} \cline{3-4} 
E23    &                                            & \multirow{4}{*}{TuYa Thermostat} & On                            \\ \cline{1-1} \cline{4-4} 
E24    &                                            &                                    & Off   \\ \cline{1-1} \cline{4-4}
E25    &                                            &                                    & Temperature INC   \\\cline{1-1}  \cline{4-4}
E26    &                                            &                                    & Temperature DEC   \\ \cline{1-1}\cline{3-4}
E27    &                                            & \multirow{5}{*}{HW Thermostat} & On                            \\ \cline{1-1} \cline{4-4} 
E28    &                                            &                                    & Off   \\ \cline{1-1} \cline{4-4}
E29    &                                            &                                    & Temperature INC/DEC   \\ \cline{1-1} \cline{4-4}
E30    &                                            &                                    & Mode Comfort   \\ \cline{1-1} \cline{4-4}
E31    &                                            &                                    & Mode Non-Frozen   \\ \cline{1-1}\cline{3-4}
E32    &                                            & \multirow{2}{*}{Ring Alarm} & Mute                            \\ \cline{1-1} \cline{4-4} 
E33    &                                            &                                    & Ring   \\ \hline 
\end{tabular}
\caption{Event and the corresponding IDs. ``H'' indicates events with large traffic volume.}
\label{tab:events}
\end{table*}
\section{Discussion}
\label{sec:discussion}
\noindent  \textbf{Mitigate Wireless Monitoring.} While the implementation of defenses relies on manufacturers, some smart home usage habits could help prevent attackers from conducting malicious activities. For instance, keep device firmware up to date to remove software vulnerabilities and set necessary automation for security-related devices to ensure they are in the expected states. Conversely, manufacturers need to actively implement certain anonymization methods to cover the trails of devices.

\noindent  \textbf{Limitations of WiFinger.} Although WiFinger achieves satisfactory accuracy and provides options for parameter adaptation, it does not fully address the problem of packet loss. To alleviate the impact, we plan to explore the use of traffic recovery techniques, e.g., data synthesis~\cite{jiang2024netdiffusion,chu2025netssm}, in the future. In addition, the current WiFinger still requires pre-collecting offline training data before deploying the attack. A more ideal scenario is that attackers could directly sniff traffic from various devices and leverage contextual information such as placement position, signal strength, and activation time to infer potential events using unlabeled data.

\noindent \textbf{Streaming IoT Devices/Events.} Streaming IoT devices such as Smart Cameras belong to another dominant class. When turned on, these devices constantly stream video data to the cloud using the UDP protocol so that users can view the live feed via the companion app. However, when triggering commands such as ``photo capture'' or ``video recording'', some tested devices (e.g., Wyze Camera, XiaoMi Camera) did not demonstrate characteristic WiFi traffic patterns. Diving into their TCP/IP traffic~\cite{ren-imc19}, we notice that no TCP packet was transmitted during events, indicating that ``photo capture'' and ``video recording'' for some cameras are likely local operations where the mobile device takes a snapshot of the video stream. Therefore, these events are simply not observable from the network traffic perspective and cannot be fingerprinted at all. Nonetheless, most streaming devices demonstrate obvious traffic volume differences that can be utilized to determine their on/off states or ``motion detected'' events~\cite{apthorpe2017spying}.

\noindent \textbf{Ground Truth Reference for Fingerprint Extraction.} During fingerprint extraction, the incomplete training sequence poses challenges. On the other hand, TCP/IP layer traffic mirroring are usually more stable and complete. Therefore, by integrating higher-layer traffic, lower-layer missing packets might be recovered by comparing their sizes to higher-layer packet sizes. Nonetheless, one challenge lies in the packet aggregation mechanisms (e.g., A-MPDU in 802.11n and later), which can disrupt the one-to-one mapping between WiFi frames and higher-layer packets.

\noindent \textbf{Potentials Beyond WiFi and Smart Home.}
WiFinger has the potential to be applied to other non-invasive monitoring scenarios as well. For example, fingerprinting medical devices such as blood glucose meters and heart rate monitors provides multi-modal information for detecting anomaly behaviors~\cite{mashnoor2024network}, e.g., uneven heartbeat. In the industrial domain, WiFinger may be used to fingerprint smart meters or smart sensors and help detect anomalies in the power grid or manufacturing processes. Apart from IoT systems, WiFinger also demonstrates its potential of fingerprinting complex network events such as mobile application behaviors, being useful for unveiling privacy leaking activities in the background~\cite{van2020flowprint,jiang2022accurate}. 

\noindent \textbf{Other security and privacy issues of Wireless IoT systems.}
Location privacy concerns in wireless IoT systems has been a long-standing problem. Prior works have demonstrated that traffic information from various layers can be leveraged to reveal users' locations~\cite{li2024rftrack,zheng2019missile}. Meanwhile, efforts have been made to guarantee security and privacy in wireless sensing networks~\cite{chen2015privacy,hu2017assuring}.

\section{Ethical Considerations}
This research strictly adheres to ethical guidelines for academic study. The nature of this work is fundamentally defensive, with the primary aim of exposing vulnerabilities to inform the development of stronger security countermeasures. All experiments were conducted within a controlled laboratory environment using devices owned by the research team. This methodology prevents any content leakage, guarantees that no personally identifiable information is exposed, and ensures no involuntary participation occurred during the wireless data collection phase. Furthermore, the dataset to be released has undergone a meticulous cleansing and anonymization process. This process removes any extraneous or potentially sensitive traffic, ensuring full compliance with data privacy standards.

\section{Conclusion}\label{sec:conclusion}
In this work, we proposed WiFinger to extract the packet-level event fingerprints from noisy wireless traffic. We demonstrated that the current trend of using ML approaches for fingerprinting IoT events has inherent overheads and limitations, especially when applied to Wi-Fi. Additionally, we identified a gap in existing evaluation methodologies, which often use chunked traffic samples rather than the more appropriate approach of detecting events within continuous traffic streams. Our experiments show that WiFinger achieves the best performance under more practical settings, maintaining very low false positive rates.



\bibliographystyle{IEEEtran}
\bibliography{paper}

\appendix
\subsection{NP-hard Proof}
\label{sec:app:np_hard}
To prove the NP-hardness of the problem, we reduce the \textit{Maximum Cardinality Subset} (MCS) problem to the \textit{FMLCS with time constraints} problem. The \textit{Maximum Cardinality Subset} problem is defined as follows. Giving a non negative set, 
\[S=\{a_1, a_2, a_3, \dots, a_n\}\]
the target is to find a subset $S' \subseteq S$ with the most elements such that the sum of the elements in $S'$ does not exceed threshold $K$. The \textit{Maximum Cardinality Subset} problem has already been proven to be NP-complete.

\subsubsection{Reduction}
We reduce the MCS problem to the FMLCS problem. For the FMLCS problem, two sequences are constructed:
\begin{equation}
    \begin{aligned}
        Seq_1 = [&\{time:\sqrt{a_1}, size:1, dir: 1\},\\
                &\{time:\sqrt{a_2}, size:1, dir: 1\},\\
                &\{time:\sqrt{a_3}, size:1, dir: 1\},\dots, \\
                &\{time:\sqrt{a_n}, size:1, dir: 1\}], \\
                \quad a_i \in S
    \end{aligned}
\end{equation}

\begin{equation}
    \begin{aligned}
        Seq_2 = [&\{time:0, size:1, dir: 1\},\\
                &\{time:0, size:1, dir: 1\},\\
                &\{time:0, size:1, dir: 1\}, \dots, \\
                &\{time:0, size:1, dir: 1\}]
    \end{aligned}
\end{equation}
For each element $a_i \in S$, we construct a sequence $Seq_1$ with each element using $a_i$ as the timestamp, and another sequence $Seq_2$ with all elements having $time=0$. As the elements in both $Seq$ have the same size, the only constraint is that the L2Norm distance between timestamps of the selected subsequence should be lower than $\sqrt{K}$. 

\subsubsection{Equivalence}
If there exists a subset $S' \subseteq S$ such that the sum of the elements in $S'$ does not exceed threshold $K$, then the FMLCS problem has a solution. The solution is to select the elements in $Seq_1$ that correspond to the elements in $S'$, and the L2Norm distance between the timestamps of the selected elements is less than $\sqrt{K}$.
\begin{equation}
    \begin{aligned}
        \sum_{a_i \in S'} a_i \leq K \Rightarrow \sum_{a_i \in S'} (\sqrt{a_i}-0)^2 \leq (\sqrt{K})^2
    \end{aligned}
\end{equation}
Therefore, since MCS is NP-Complete, FMLCS has a polynomial time solution only if P=NP.

\end{document}